\documentclass[prx,superscriptaddress,aps,twocolumn]{revtex4-2}
\usepackage{amsmath}  
\usepackage{amsthm}  
\usepackage{float}
\usepackage{comment}
\usepackage{soul}
\usepackage{amssymb}	
\usepackage{wasysym} 
\usepackage{soul}
\usepackage{longtable}
\usepackage{graphicx}  
\usepackage[usenames,dvipsnames]{color}
\usepackage{array} 
\usepackage{paralist} 
\usepackage{amsfonts}
\usepackage{mathtools}
\usepackage{times}
\usepackage{braket}
\usepackage{slashed}
\usepackage{multirow}
\usepackage[colorlinks=true,linkcolor=blue,citecolor=blue,breaklinks]{hyperref}
\usepackage{cleveref}

\crefformat{figure}{#2{\color{Maroon}#1}#3}
\normalfont


\renewcommand{\vec}[1]{\ensuremath{\mathbf{#1}}} 
\newcommand{\avg}[1]{\left< #1 \right>} 
\let\baraccent=\= 
\renewcommand{\=}[1]{\stackrel{#1}{=}} 

\newcommand{\be}{\begin{equation}}
\newcommand{\ee}{\end{equation}}

\newcommand{\bea}{\begin{eqnarray}}
\newcommand{\eea}{\end{eqnarray}}
\newcommand{\beal}{\begin{align}}
\newcommand{\eeal}{\end{align}}

\newcommand{\pdg}{{\phantom\dagger}}

\begin{document}


\title{Elementary Building Blocks for Cluster Mott Insulators}

\author{Vaishnavi Jayakumar}
\email[E-mail: ]{vj@thp.uni-koeln.de}
\affiliation{Institute for Theoretical Physics, University of Cologne, 50937 Cologne, Germany}
\author{Ciar\'{a}n Hickey}
\email[E-mail: ]{ciaran.hickey@ucd.ie}
\affiliation{Institute for Theoretical Physics, University of Cologne, 50937 Cologne, Germany}
\affiliation{School of Physics, University College Dublin, Belfield, Dublin 4, Ireland}
\affiliation{Centre for Quantum Engineering, Science, and Technology, University College Dublin, Dublin 4, Ireland}

\begin{abstract}
Mott insulators, in which strong Coulomb interactions fully localize electrons on single atomic sites, play host to an incredibly rich and exciting array of strongly correlated physics. One can naturally extend this concept to cluster Mott insulators, wherein electrons localize not on single atoms but across clusters of atoms, forming ``molecules in solids''. The resulting localized degrees of freedom incorporate the full spectrum of electronic degrees of freedom, spin, orbital, and charge. These serve as the building blocks for cluster Mott insulators, and understanding them is an important first step toward understanding the many-body physics that emerges in candidate cluster Mott insulator materials. Here, we focus on elementary building blocks, neglecting some of the complexity present in real materials that can often obfuscate the underlying principles at play. Through an extensive set of exact theoretical calculations on clusters of varying geometry, number of orbitals, and number of electrons, we uncover some of the basic organizing principles of cluster Mott degrees of freedom, particularly when interactions dominate and negate a simple single-particle picture. These include, for example, the identification of an additional ``cluster Hund's rule'', of cluster ground states that are best understood from a purely interacting perspective, and of several localized degrees of freedom which are protected by an unusual combination of discrete spatial or orbital symmetries. Finally, we discuss the impact of adding additional terms, relevant to material candidates, on the phase diagrams presented throughout, as well as the potential next steps in the journey to building a more complete picture of cluster Mott insulators. 

\end{abstract}

\maketitle

Strong electronic interactions can drive the emergence of new and distinct quantum phases of matter. The Mott insulating phase serves as one of the most striking and well-studied examples, in which strong on-site Coulomb interactions fully localize electrons on single atomic sites. In the simplest single-orbital description of such a scenario, the resulting localized degrees of freedom carry spin-$1/2$ and interact via an effective spin Hamiltonian due to virtual exchange processes. Such \emph{conventional} Mott insulators play host to an incredibly rich and complex array of quantum magnetism \cite{sachdev_mott,qmagnetism_milestones,qsl_review}, proving relevant to a wide variety of different materials \cite{cuprate_rmp,herbertsmithite,organic_mott}.  

Beyond the simplest single-orbital scenario, the addition of multiple orbitals introduces new physics and extends the possibilities for Mott physics. The central idea, that strong on-site Coulomb interactions fully localize electrons on single atomic sites, remains. However, with those localized electrons now carrying both spin and orbital character, the resulting localized degrees of freedom can be purely spin, purely orbital, or a combination, depending on the interplay between intra-orbital interactions, spin-orbit coupling, and crystal field splittings. A prominent example are the \emph{spin-orbit entangled} Mott insulators, including the $j_{\textrm{eff}}=1/2$ family of compounds, in which the local moments are a combination of both spin and orbital degrees of freedom \cite{bjkim}. This has the knock-on effect of dramatically altering the nature of the allowed effective spin Hamiltonians, with, for example, new bond-dependent interactions such as the Kitaev interaction potentially possible \cite{mott_jackeli, kitaevmaterials_ciaran, winter_kitaev}. 

The final natural extension of the above ideas is to expand the localized degrees of freedom to include, as well as the spin and orbital degrees of freedom, the charge degree of freedom of the electron. If the lattice hosts well-defined clusters of atoms, e.g.~dimers or trimers, then strong interactions, coupled with an imbalance of inter and intra-cluster hopping, can drive electrons into localizing on these clusters, rather than on single atomic sites. In this scenario, the electrons are delocalized across each cluster, but localized between clusters such that the system remains a charge insulator. The localized degrees of freedom can be naturally described in terms of molecular orbitals, giving rise to what are sometimes referred to as ``molecules in solids" \cite{khomskii_review}, and can now be a combination of different spin, orbital, and positions of the electron. Such \emph{cluster} Mott insulators (CMIs) have become increasingly prevalent in recent years \cite{Cava, Cava_structural,m2o9dimer}. 
The clusters can either be built into the crystal structure from the start, or they can spontaneously emerge via cooperative structural transitions at finite temperature, as for example in titanium pyroxenes \cite{pyroxene}, LiVO$_2$ \cite{LiVO2}, or AlV$_2$O$_4$ \cite{heptamers}. One can similarly imagine cluster analogues of generalized Wigner crystals \cite{Wigner1978}, in which the interaction-driven localization of electrons on clusters of sites spontaneously breaks lattice symmetries.
The clusters themselves can range from simple dimers, such as in the M$_2$O$_9$ family of compounds \cite{RIXS_M2O9_disorder,RIXS_Ba3InIr2O9,  Qiang_Ba3LaRu2O9, Qiang_Momagnets,Qiang_Ba3CeRu2O9,chargeorder_Ba3NaRu2O9, chargeorder_Ba3NaRu2O9,Ba3ZnIr2O9},  Ba$_5$AlIr$_2$O$_{11}$\cite{Khomskii_SOCandHund}, or Ba$_5$Mn$_3$O$_{12}$Cl \cite{Cava_dimerkagome}, to triangles as in the Mo$_3$O$_8$ family of compounds \cite{nikolaev,Gangchen_kagome,Gangchen_CMI_CurieWeiss,honeycomb_LiZn2Mo3O8,Gangchen_clustertransition} or niobium halides \cite{nb3cl8_flatbands,Nb3Cl8_phasetrans,Nb3X8_triangular_clusters,clustermagnet_vanderwaals}, and even pyrochlore lattice systems \cite{gangchen_pyrochlore} with triangular clusters as in CsW$_2$O$_6$ \cite{triangle_CsW2O6}, or with tetrahedral clusters, as in the lacunar spinel compound GaTa$_4$Se$_8$ \cite{GaTa4Se8,spinel_bondorder,lacunarspinel_sc}. Finally, the physics at play ranges from potential valence bond condensation in Li$_3$Zn$_2$Mo$_3$O$_8$ \cite{Mo3o8_valencebond}, to candidate quantum spin liquids in Ba$_3$InIr$_2$O$_9$ \cite{RIXS_Ba3InIr2O9}, Ba$_4$Nb$_{0.8}$Ir$_{3.2}$O$_{12}$ \cite{Ba4spinliq}, or Na$_3$Sc$_2$Mo$_5$O$_{16}$\cite{Qiang_Momagnets}, and even includes a purported unusual combination of heavy-fermion, strange metal and spin liquid physics in the Ba$_4$Nb$_{1-x}$Ru$_{3+x}$O$_{12}$ series of compounds \cite{heavyfermiontrimer}. 

The basic building blocks of CMIs are of course the clusters themselves and their associated localized degrees of freedom. Thus, the first step on the path to a comprehensive understanding of CMIs is to understand the physics of individual clusters and how and which localized degrees of freedom can emerge. However, in real materials, there are a multitude of factors that determine the localized cluster degrees of freedom, e.g.~the local symmetry of the cluster, number of electrons, nature and strength of the intra-cluster hopping, crystal field splitting, spin-orbit coupling, Hubbard $U$ and Hund's coupling $J$, and more. This extremely high-dimensional parameter space can make it difficult to obtain a clear picture of overarching organizing principles, particularly when interactions are relevant and negate a simple single-particle picture. Rather, one is typically forced to address the issue on an individual case-by-case basis.  

Here, we study a set of simple, elementary building blocks for CMIs, revealing a number of insights into how non-trivial localized degrees of freedom can emerge. We achieve this by restricting the number of terms that we consider in our cluster Hamiltonians, and neglecting the spatial characteristics of the orbitals, while maintaining the core ingredients of cluster physics, namely (i) a molecular orbital level structure labelled by irreducible representations of the cluster point group and, in some cases, discrete internal orbital symmetries, and (ii) on-site Hubbard-Kanamori interactions. We make the further simplifying assumption that spin $SU(2)$ symmetry is preserved (i.e. that spin-orbit coupling
is negligible). The advantage of such an approach is that it makes the parameter space manageable and, more importantly, makes it easier to disentangle which terms are responsible for which outcomes. The disadvantage of course is that, by restricting the number of allowed terms, we are not able to capture the full complexity of real materials. However, at the end, in Section \ref{discussion}, we will discuss the consequences of adding more realistic terms and how they would impact the results. More concretely, we study a variety of cluster geometries, including dimers, trimers, triangles, squares, tetramers, and tetrahedra, as well as cases with one, two and three orbitals per site, and all possible electron fillings. For each, we use exact diagonalization to find the ground state(s) of the relevant simplified cluster Hamiltonian as one varies the interaction strengths of on-site Hubbard $U$ and Hund's coupling $J$. We present only a subset of the data here that best illustrates the key physics, with the remainder of the data and phase diagrams available in a public repository \cite{cmi_data}.      

The three key insights that can be gleaned from our extensive set of calculations can be summarized as follows: (i) Focusing on the interacting part of the cluster Hamiltonian, which, for each site, we take as the standard multi-orbital Hubbard-Kanamori Hamiltonian, and expressing it in terms of $N_i^2$, $\vec{S}_i^2$ and the orbital operators, e.g.~$\vec{L}_i^2$ in the three-orbital case, we find that there is an additional ``cluster Hund's rule" that must be obeyed. First, $\sum_i N_i^2$ must be minimized, before then applying the cluster versions of the usual Hund's rules, maximal $\sum_i \vec{S}_i^2$ and then maximal $\sum_i\vec{L}_i^2$ (in the three-orbital case). However, depending on the ratio of $J/U$, a more complex competition may arise, which negates a straightforward set of cluster Hund's rules. Such competition is similar to the physics that can lead to Hund's insulators and mixed valence states in lattice systems \cite{Isidori2019,Ryee2021}. (ii) While some ground states can be easily understood starting from the non-interacting Hamiltonian and its single-particle molecular orbital levels, there are other ground states that can only be understood starting from the purely interacting Hamiltonian. They do not smoothly connect to any non-interacting limit. (iii) A majority of parameter space, at least explored in this study, largely contains either a unique ground state in the case of an even number of electrons, or a degeneracy solely due to spin in the case of an odd number of electrons. Non-trivial degeneracies protected either by lattice or orbital symmetries, though possible, are rare.  

The paper is structured as follows: Section \ref{model_methods} introduces the general setup, Hamiltonian and methods used throughout. Sections \ref{single_orb}, \ref{2orb} and \ref{3orb} present the cases of cluster Mott insulators with one, two, and three orbitals per site respectively, providing in each case a select number of illustrative phase diagrams for different cluster geometries and electron fillings. Finally, we end, in section \ref{discussion}, with a discussion of the key insights and implications of our work for future studies, as well as the impact of adding more terms relevant to candidate materials.

\section{Model and Methods} \label{model_methods}

As a useful comparison, let us first consider the Hamiltonian for the simple single-band Hubbard model

\begin{equation}
    H = + U \sum_{i} n_{i\uparrow}n_{i\downarrow} -t\sum_{\langle i,j \rangle, \sigma} (c^\dagger_{i\sigma}c_{j\sigma} + h.c) ,
    \label{eqn:hubbard}
\end{equation}
where $c^\dagger_{i\sigma}(c_{i\sigma})$ creates (annihilates) an electron with spin $\sigma \in \{\uparrow,\downarrow\}$ on site $i$. 

\begin{figure}
    \includegraphics[width=6.5cm,height=2.5cm]{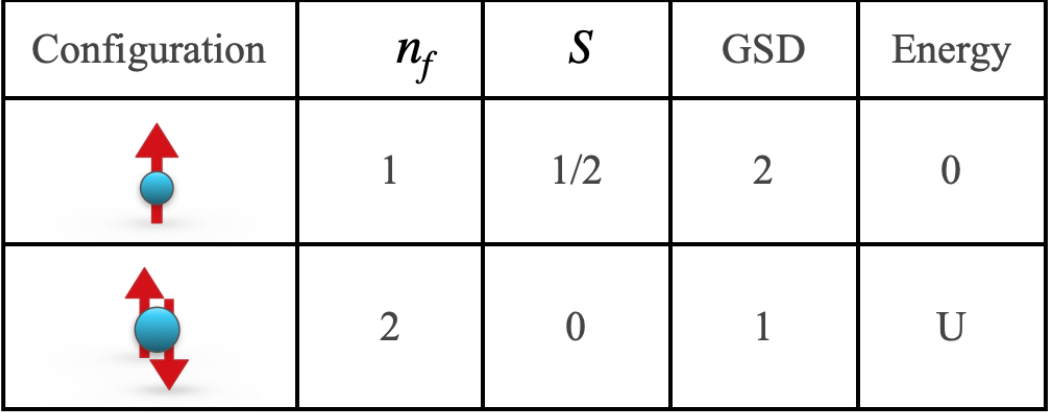}
    \caption{Spectrum of the Hubbard model for a single site. A singly occupied, that is, half-filled site is of lower energy (i.e, zero) than a doubly occupied one, which comes with a huge cost of $U$ in the strongly interacting limit. Note that the trivial case of $n_f=0$ also has zero energy.}
    \label{fig:hubbard_single_site}
\end{figure}

The Mott insulating phase is obtained at half-filling, i.e.~one electron per site, when $U\gg t$, that is, in the strong interaction limit of the Hubbard Model. In this limit, the Hamiltonian can be split as
\begin{equation}
    H = H_0 + V,
    \label{eqn:intracluster_h0v}
\end{equation}
where $H_0$ is the on-site Hubbard term and the hopping term between sites is treated as a perturbation $V$. Using this as a starting point, we proceed in two steps. (i) First, we need to find the possible ground states of $H_0$. Since it is a sum of single-site terms, it suffices to study just a single site. Fig.~\cref{fig:hubbard_single_site} indicates the possible electron fillings $n_f$ for a single site, and the corresponding energies of $H_0$. While a singly-occupied site has zero energy, adding an extra electron generates a large energy penalty of $U$, making the singly-occupied state the lower energy state. Extending this to the full lattice, we find that the ground states at half-filling consist of all states with precisely one electron per site. Thus, there is a localized effective $S=1/2$ degree of freedom at each site. (ii) Second, once the ground states of $H_0$ are determined, one can derive an effective Heisenberg model perturbatively in the hopping $V$, which correctly captures the magnetic physics of the Mott insulating phase.

Now, we can straightforwardly extend the above idea to a ``cluster Hubbard" Hamiltonian with a general form given by
\begin{equation}
H =  \sum_C H_C + \sum_{\langle C,C'\rangle}H_{CC'}
\label{eqn:chubbard}
\end{equation}
where $H_C$ is the intra-cluster Hamiltonian, containing electronic interactions and hopping between sites within the cluster $C$, and $H_{CC'}$ is the intra-cluster Hamiltonian, containing interactions and hopping between sites belonging to neighboring clusters $C$ and $C'$. 

If the intra-cluster terms dominate, we can again split the Hamiltonian into an unperturbed Hamiltonian $H_0=\sum_C H_C$, which is nothing but the intra-cluster Hamiltonian summed over all clusters, and then treat the inter-cluster terms as perturbations, $V=\sum_{CC'} H_{CC'}$. The way forward again proceeds with the same two steps as in the more familiar single site, single orbital case discussed above. (i) First, we need to find the possible ground states of $H_0$. Since this is a simple sum over clusters, it suffices to study just a single cluster. The ground states determine the potential localized degrees of freedom available. (ii) Once the ground states of $H_0$ are determined, an effective Hamiltonian can be derived perturbatively in $V$, which links the clusters together, assuming that the chosen electron filling results in a degenerate set of ground states. 

In this work, we focus exclusively on the first step mentioned above: identifying the possible ground states of the intra-cluster Hamiltonian and, consequently, the various potential localized degrees of freedom.

\subsection{Intra-Cluster Hopping and Molecular Orbitals} \label{model_methods_nonint}

\begin{figure}
    \centering
    \includegraphics[width=8cm,height=6.4cm]{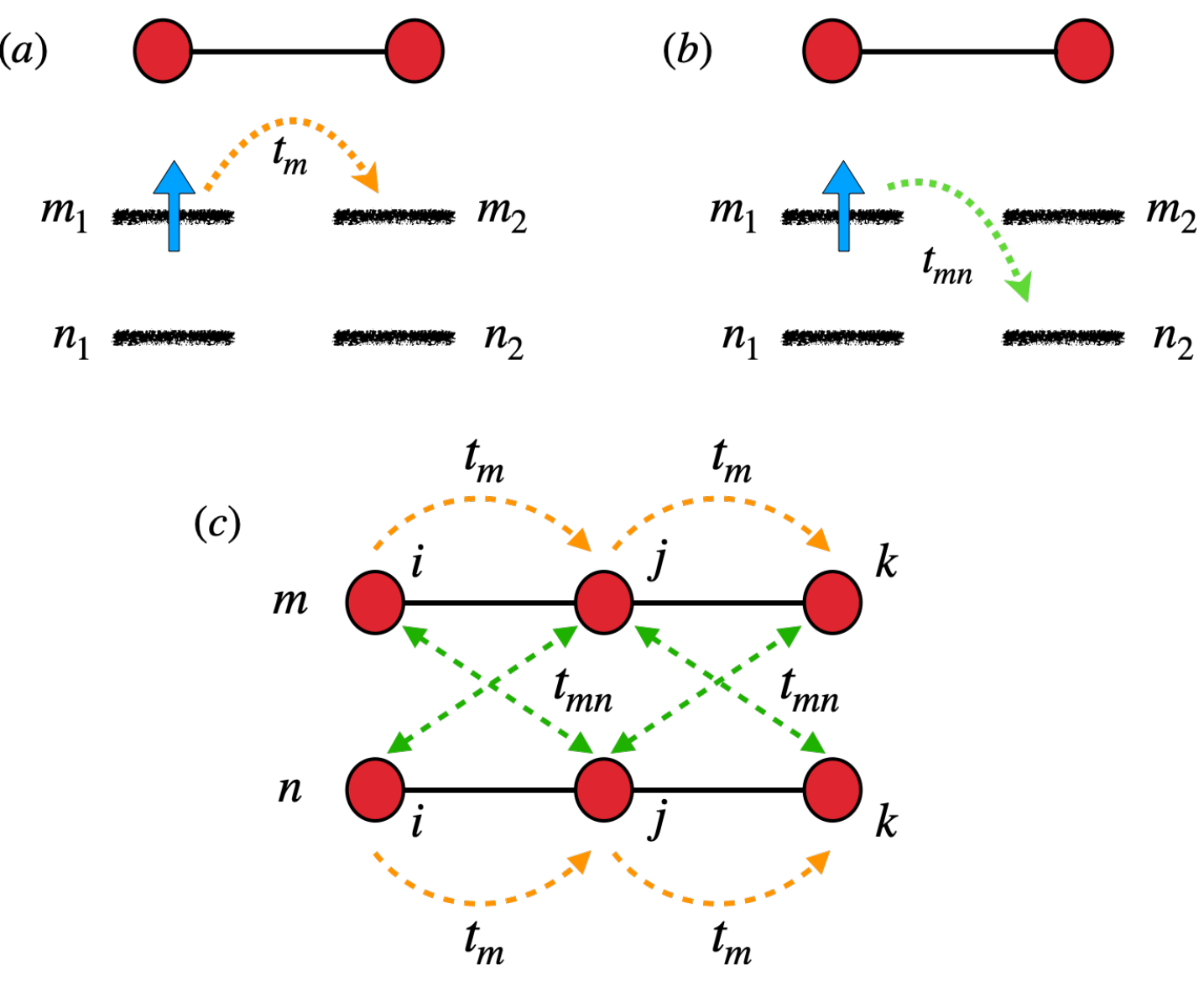}
    \caption{Hopping mechanisms in $H_{\textrm{non-int}}$. (a) $t_m$ hops an electron from one orbital on site $i$ to the same kind of orbital on site $j$. (b) $t_{mn}$ hops an electron from one orbital on site $i$ to a different kind of orbital on site $j$. (c) Another way of understanding the two kinds of hopping is that $t_m$ (shown in orange) hops spins in the same ``orbital plane" whereas $t_{mn}$ hops spins across different orbital planes (shown in green).}
    \label{fig:hopping_mech}
\end{figure}

The non-interacting part of the intra-cluster Hamiltonians we consider in this study contains only hopping terms. The potential impact of other non-interacting terms relevant to many materials, such as crystal-field splitting or spin-orbit coupling, will be discussed in Section \ref{discussion}. For the single orbital case, we consider the simplest intra-cluster hopping Hamiltonian
\begin{equation}
H_{\textrm{non-int}} = -t\sum_{\langle i,j \rangle, \sigma} (c^\dagger_{i\sigma}c_{j\sigma} + h.c).
\label{eqn:ordinary_hopping}
\end{equation}
For the multi-orbital case, in real materials, the hopping Hamiltonian can be constructed using knowledge of the orbitals involved (including surrounding ligands), the point-group symmetry of the cluster, and the relevant Slater-Koster parameters. Here, as we are aiming for a simpler, more overarching perspective, we consider a simplified form of intra-cluster hopping as
\begin{align}
 \notag   H_{\textrm{non-int}}  = &-t_{m}\sum_{\langle i,j\rangle} \sum_{m, \sigma} (c^\dagger_{ im\sigma}c^\pdg_{ jm\sigma} + h.c) \\
    & -t_{mn}\sum_{\langle i,j\rangle} \sum_{m\neq n,\sigma} (c^\dagger_{ im\sigma}c^\pdg_{ jn\sigma} + h.c) ,
    \label{eqn:multiorb_hopping_gen}
\end{align}
where $t_m$ corresponds to diagonal, intra-orbital hopping, $t_{mn}$ to off-diagonal, inter-orbital hopping and the operator $c^\dagger_{im\sigma}(c_{im\sigma})$ creates (annihilates) an electron with spin $\sigma$ in atomic orbital $m$ on site $i$. This form of hopping is illustrated in Fig.~\ref{fig:hopping_mech}. 

We refer to the energy levels of the non-interacting Hamiltonian as molecular orbitals. The symmetries of the molecular orbital levels have two contributions. The first contribution comes from the spatial symmetry of the cluster. As we will be agnostic regarding the spatial characteristics of the orbitals, open chains, including the dimer, trimer, and tetramer clusters are assumed to have only inversion symmetry, and hence designated as belonging to an ``$i$" point group, with $[+]$ and $[-]$ indicating even or odd under inversion respectively. The point groups of the other clusters are indicated in Fig.~\cref{fig:1orb_table}. The second contribution comes from the internal orbital symmetry among the orbitals themselves. In the absence of the off-diagonal hopping $t_{mn}$, the hopping Hamiltonian has an enlarged SU(2) and SO(3) orbital symmetry for the case of two and three orbitals respectively. However, finite $t_{mn}$ breaks these symmetries, with the two-orbital case reduced to a $C_2$ orbital point group (corresponding to swapping of the two orbitals), and the three-orbital case reduced to a $C_{3v}$ orbital point group (corresponding to cyclic permutations of the orbitals and swapping of any two). We will use a shorthand $[G_C,G_O]$ notation, with $G_C$ referring to the spatial point group of the cluster and $G_O$ referring to the orbital symmetry group. Finally, the Hamiltonian is time-reversal symmetric, meaning that all single-particle levels possess a two-fold Kramers degeneracy. 

\subsection{Intra-Cluster Interactions} \label{model_methods_int}
 
\begin{figure}
    \centering
    \includegraphics[width=8cm,height=3.3cm]{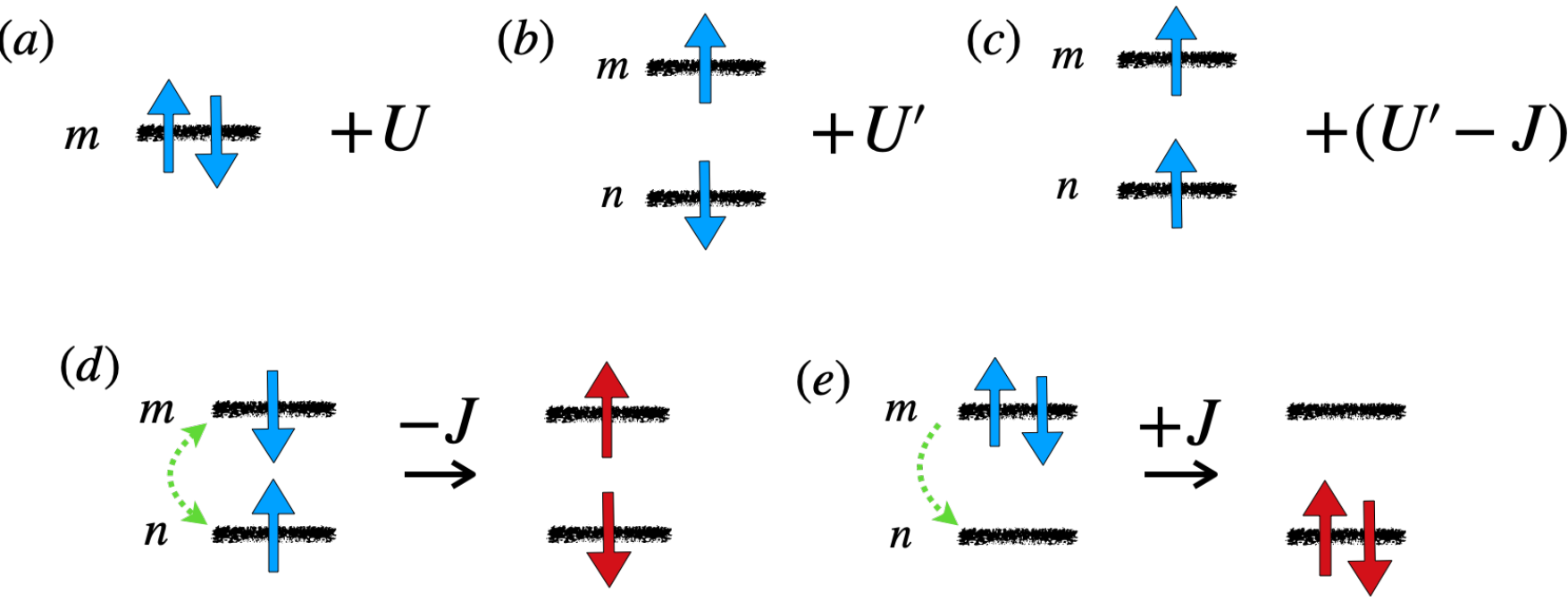}
    \caption{Nature of interactions governed by various terms of the Hubbard-Kanamori Hamiltonian. (a) The on-site $U$ term, (b) density term for opposite spins
in orbitals $m$ and $n$, (c) density term for parallel spins orbitals $m$ and $n$, (d) two opposite spins  on orbitals $m$ and $n$ are flipped at the same time, (b) a pair of spins, initially on orbital $m$, hop at the same time to orbital $n$.}
    \label{fig:int_mech}
\end{figure}

For the intra-cluster interaction Hamiltonian, we consider the standard multi-orbital Hubbard-Kanamori Hamiltonian on each site, with the resulting total Hamiltonian given by
\begin{equation}
\begin{split}
     H_{\textrm{int}}  = &\sum_i \Bigg( \Bigg. U\sum_{m}n_{im\uparrow}n_{im\downarrow}+U'\sum_{m\neq n }n_{im\uparrow}n_{in\downarrow} \\
    & +(U'-J)\sum_{m\neq n,\sigma}n_{im\sigma}n_{in\sigma}  - J \sum_{m\neq n}c^{\dagger}_{im\uparrow}c^{\dagger}_{in\downarrow}c_{im\downarrow}c_{in\uparrow}\\
 & + J \sum_{m\neq n}c^{\dagger}_{im\uparrow}c^{\dagger}_{im\downarrow}c_{in\downarrow}c_{in\uparrow}\Bigg. \Bigg).
 \end{split}
 \label{eqn:hk_site}
\end{equation}
where, from now on, we set $U'=U-2J$ (for more details, see Appendix \ref{app:HK}). The physical mechanisms corresponding to each individual term are illustrated in Fig.~\cref{fig:int_mech}. In the case of a single orbital, only the first term survives. 

It's possible to rewrite the Hubbard-Kanamori Hamiltonian in a more compact form \cite{casimir}. Defining orbital operators for the two and three orbital cases as
\begin{align}
    \textrm{Two-orbitals:}& \,\,\, T^\alpha_i = \frac{1}{2}\sum_{\sigma}\sum_{mm'}c^\dagger_{im\sigma}\tau^\alpha_{mm'}c^\pdg_{im' \sigma},\\
    \textrm{Three-orbitals:}& \,\,\, L^m_i = i\sum_{\sigma}\sum_{m'm''}\epsilon^\pdg_{mm'm''}c^\dagger_{im'\sigma}c^\pdg_{im''\sigma},
\end{align}
where $\tau^\alpha$ are the Pauli matrices, and with the spin operator similarly defined as 
\begin{equation}
    S^\alpha_i = \frac{1}{2}\sum_m \sum_{\sigma\sigma'} c^\dagger_{im\sigma}\tau^\alpha_{\sigma\sigma'}c^\pdg_{im\sigma'},
   \label{eqn:s}
\end{equation}
we can write the full interaction Hamiltonian above as
\begin{equation}
     H_{\textrm{int}}  = \frac{(U-3J)}{2}\sum_i N_i^2 - 2J\sum_i(\,\vec{S}_i^2 +Q_i^2)+ \alpha n_f,
     \label{eqn:hk_gen}
\end{equation}
where $Q_i$ is an orbital operator that depends on the number of orbitals, $Q_i=T_i^y$ and $Q_i=\vec{L}_i/2$ in the two and three-orbital cases respectively, and $\alpha n_f$ is akin to a chemical potential: $n_f$ is the number of electrons on the cluster and $\alpha$ is a constant with $\alpha = (7J-U)/2$ for the two orbital and $\alpha = (8J-U)/2$ for the three orbital case. Notably, the first term goes to zero and changes sign when $J=U/3$ (indeed $U-3J$ is sometimes referred to as an effective interaction $U_{\textrm{eff}}$ \cite{Georges_2013}). Typically, the regime $J>U/3$ is discounted, as certain Coulomb matrix elements become negative. However, there are cases, such as rare-earth nickelates \cite{Alaska2015}, where certain mechanisms have been proposed that effectively realize such a regime. We will explore the full $U,J$ parameter space in order to gain a more complete picture of the underlying physics at play in the subsequent sections. Rewriting the interaction Hamiltonian in the form of Eq.~\ref{eqn:hk_gen} provides a far more intuitive understanding of the nature of the ground states favored in different limiting cases. 

Note that, if the individual $N_i$ are fixed, as in the case of a single-site Mott insulator, then the $N_i^2$ term in Eq.~\ref{eqn:hk_gen} is a constant and hence does not play a role. In that case, states with maximal $\vec{S}_i^2$ and then maximal $Q_i^2$ are favored. This is nothing but a reflection of Hund's first two selection rules. It is evident, for example, in Fig.~\cref{fig:3orbs_table}, where the quantum numbers and energies for a single site with three orbitals are given. For $n_f=2$, for example, the state with the lowest energy is the one with maximal spin $\vec{S}_i^2$ and maximal angular momentum $\vec{L}_i^2$.

In a CMI, the individual $N_i$ are not fixed; only the total electron number $n_f=\sum_i N_i$ on the cluster is fixed. In this case, the $N_i^2$ term in Eq.~\ref{eqn:hk_gen} typically dominates the energy (the eigenvalues of $N_i^2$ are typically much larger than that of $\vec{S}_i^2$ and $Q_i^2$). As a result, one gets an additional ``cluster Hund's rule" which must first be satisfied. In the physically relevant regime of $U>3J$, one must first minimize $\sum_i N_i^2$, and then, as usual, maximize $\sum_i \vec{S}_i^2$, and finally $\sum_i Q_i^2$. On the other hand, when $U<3J$, competition between the different terms may arise, potentially resulting in a more complex selection process. Minimal $\sum_i N_i^2$ favors a uniform spread of electrons across the cluster. Conversely, maximal $\sum_i N_i^2$ favors a superposition of states with a more skewed distribution of electrons. We will discuss these points in more detail in the examples that follow, in Sections \ref{2orb_phasediag} and \ref{3orb_phasediag}. 

\subsection{Comparison with conventional Hund's rules}

\begin{figure}[h]
\includegraphics[width=\linewidth]{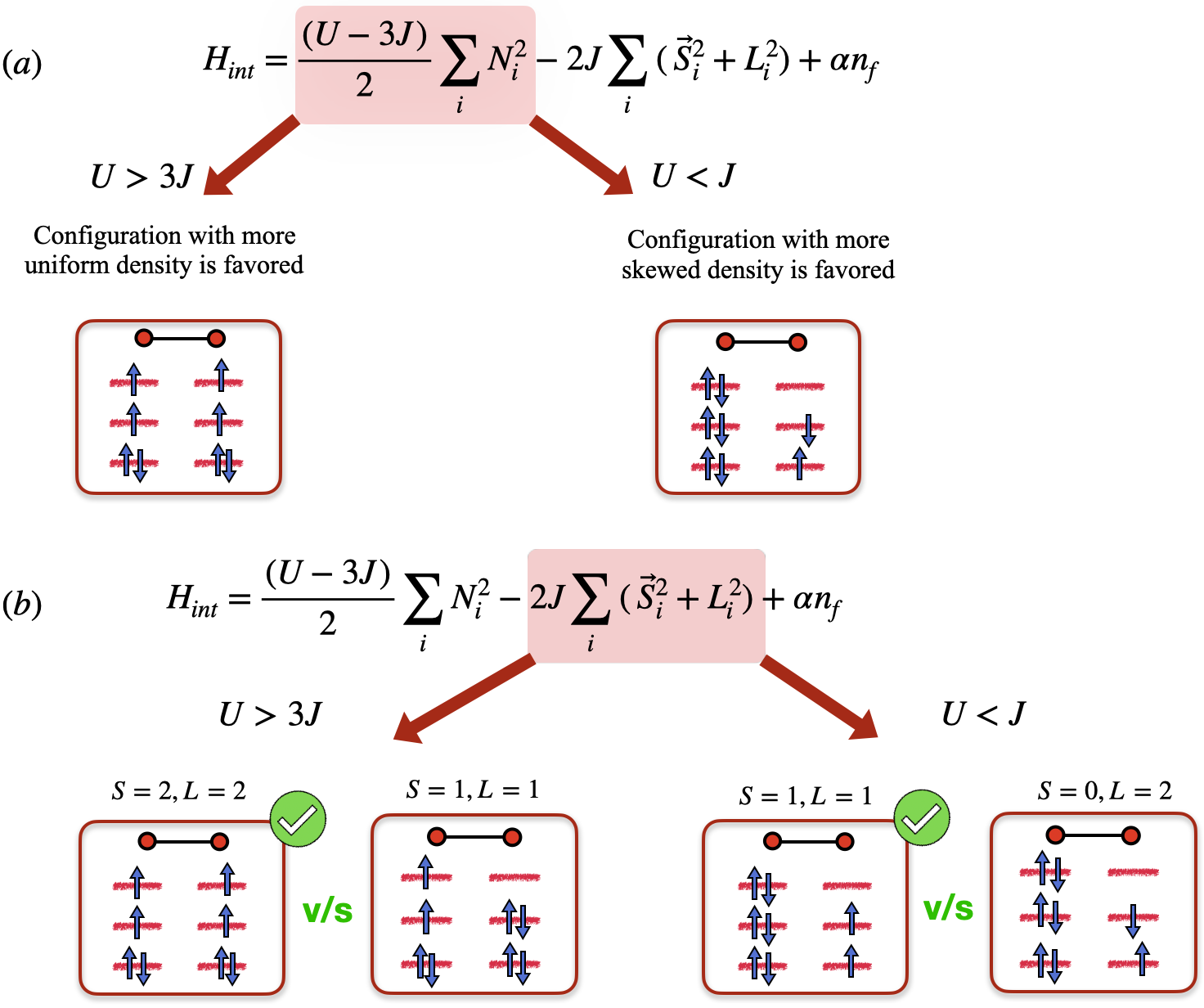}
    \caption{Example of cluster Hund’s rules at play. Here, we consider an example of a dimer cluster with 3 orbitals per site, and 8 electrons on the cluster. (a) How the first term chooses the distribution of density fluctuations in the ground state. (b) How the second term chooses the spin and angular momentum quantum numbers of the ground state. To show this, we have compared the ground state with another possible state in the respective parameter regimes. Ground states are marked with a green tick. Here, we can see that the maximum possible value of $S=2,L=2$ is attained only when we operate in the $U>3J$ regime.}
    \label{fig:clusterhundsrule}
\end{figure}

A natural question might arise at this point as to how the expectations from the additional cluster Hund's rule of the previous subsection differ from those of the conventional Hund's rules. We explore this with the help of an illustrative example, as shown in Fig.~\cref{fig:clusterhundsrule}, where we consider the case of a dimer cluster with three orbitals per site and 8 electrons on the cluster.

Let us first consider the $U>3J$ regime in Fig.~\cref{fig:clusterhundsrule}. Minimizing the $\sum_i N_i^2$ term favors a uniform density across the two sites, as shown in Fig.~\cref{fig:clusterhundsrule}(a). Given this density distribution, the second term in Eq.~\ref{eqn:hk_gen} then determines the spin and angular momentum quantum numbers. In this case, we see that the ground state has maximal $S=2,L=2$, in accordance with the usual Hund's rules. The $U>3J$ regime, in which density fluctuations are minimized, thus follows the expectations from the conventional Hund's rules. 

However, in the $U<3J$ regime, the $\sum_i N_i^2$ term must be maximized, rather than minimized. This favors a more skewed distribution of density fluctuations (it should be noted that the density itself, $\braket{N_i}$, remains uniform). Given this, the second term in Eq.~\ref{eqn:hk_gen} then selects an $S=1,L=1$ ground state, in contrast to what we would have naively expected from simply applying the conventional Hund's rules, i.e.~disregarding the $\sum_i N_i^2$ term in the interaction Hamiltonian of Eq.~\ref{eqn:hk_gen} and just maximizing the spin and angular momentum. In general, the interplay between the first and second terms of Eq.~\ref{eqn:hk_gen} becomes more intricate outside of the $U>3J$ regime, resulting in ground states where neither the spin nor angular momentum is necessarily maximized. In fact, as we will see in certain cases that will follow (see in particular section \ref{3orb_phasediag}), there can even exist an intermediate region between the $U\gg 3J$ and $U\ll3J$ limits where a different ground state to either limit is favored.

\subsection{Methods}

We solve the single-cluster Hamiltonian $H_C = H_{\textrm{int}}+H_{\textrm{non-int}}$, with $H_{\textrm{int}}$ given by Eq.~\ref{eqn:hk_site} and $H_{\textrm{non-int}}$ given by Eq.~\ref{eqn:multiorb_hopping_gen}, using exact diagonalization. For the general multi-orbital case, the Hilbert space size is $4^{N n_{\textrm{orb}}}$, where $N$ is the number of sites and $n_\textrm{orb}$ is the number of orbitals. In the absence of interactions, the non-interacting cluster Hamiltonian is easily solved as it's sufficient to diagonalize only in the single-particle sector. On the other hand, in the presence of interactions, it's necessary to consider the full Hilbert space for each $n_f$ block of the Hamiltonian. Since it is the interactions, $U$ and $J$, which render the calculation non-trivial, we thus present all phase diagrams as a function of $U$ and $J$, with each phase diagram consisting of $100\times100$ parameter points. It's important to keep in mind though that the physically relevant regime for most materials corresponds to $U>3J$. Including $U<3J$ in our phase diagrams gives us a better picture of the competition between the different ordering tendencies.

\section{Case-I: Single Orbital Per site} \label{single_orb}

As an illustrative starting point, we first discuss the case with just a single orbital per site. 

\begin{figure}
    \centering
    \includegraphics[width=7cm,height=5cm]{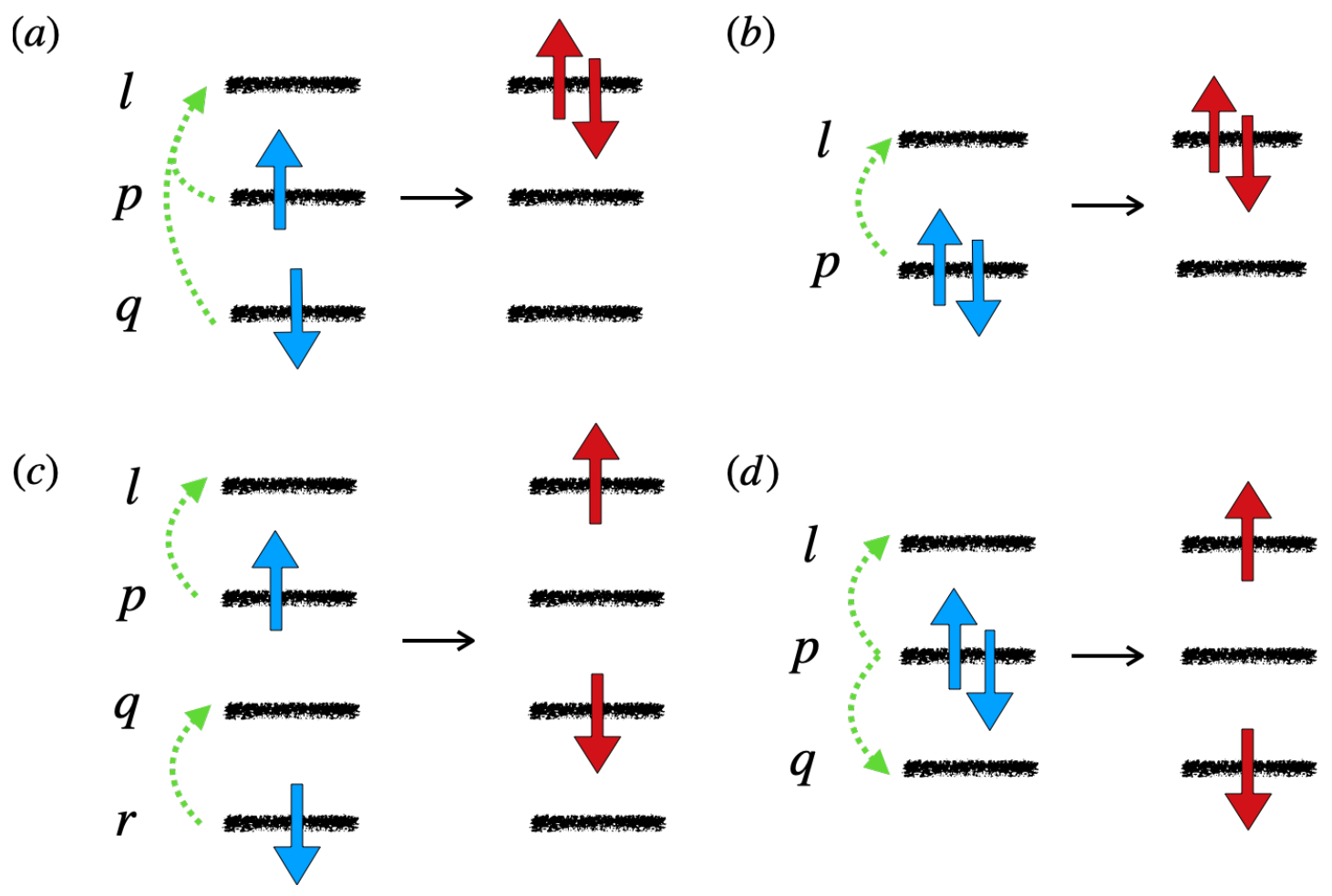}
    \caption{Various hopping mechanisms in $H_{int}^i$ for a single orbital per site. (a) pair-clumping, (b) pair-clumping becomes pair-hopping when $p=2mn-n$. (c) Two spins hop simultaneously from two orbitals to two other orbitals. (d) This becomes pair-spreading when $q=m+p-n$.}
    \label{fig:1orb_mech}
\end{figure}

\begin{figure*}[t]
    \centering
    \includegraphics[width=14.5cm,height=9.5cm]{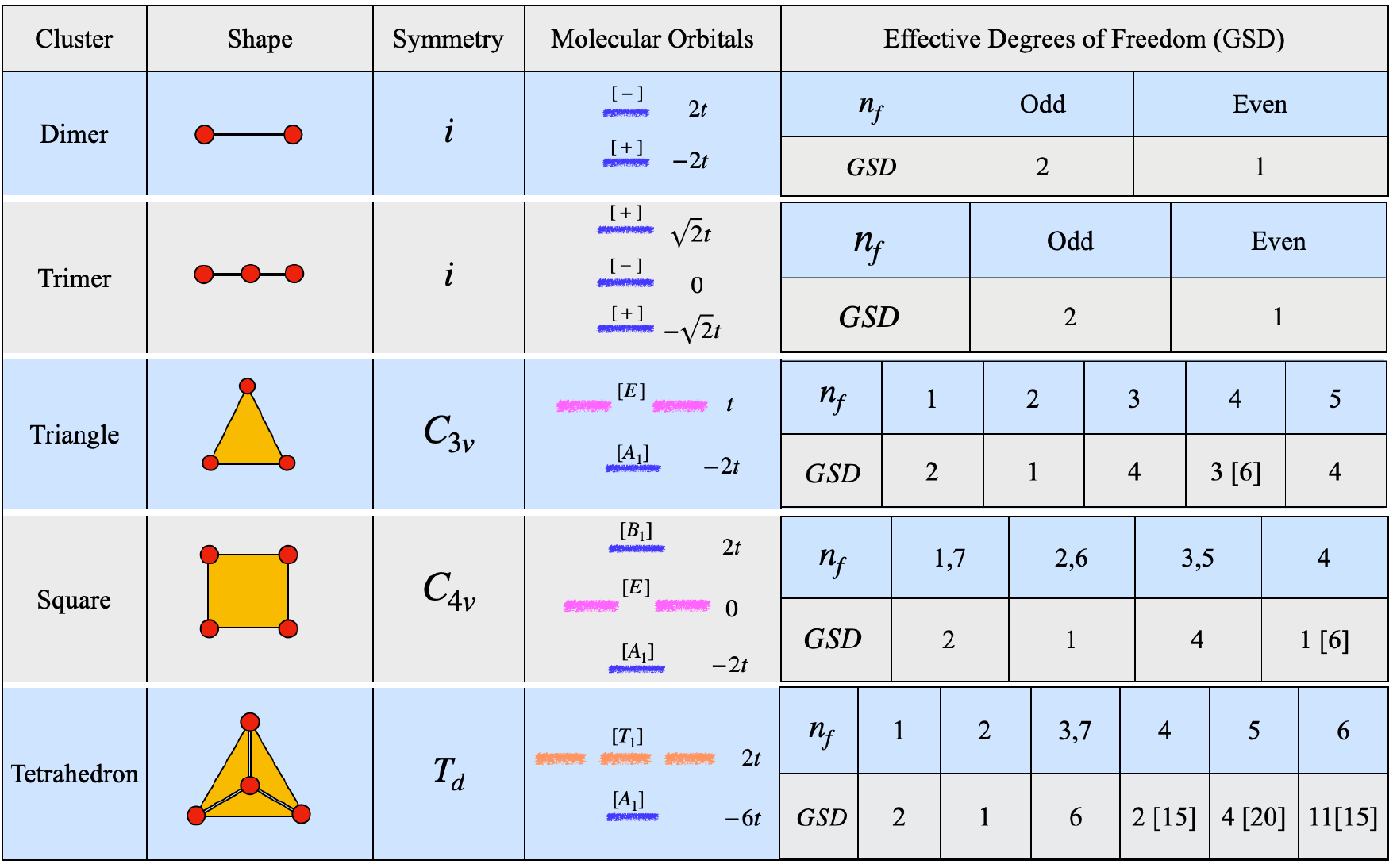}
    \caption{This figure shows the non-interacting molecular orbital levels and the GSDs for the full Hamiltonian for all electron fillings for clusters with 2,3 and 4 sites and a single orbital per site. The numbers in square brackets in the GSD row indicate the GSD in the absence of interaction $U$. When square brackets are not indicated, it means the GSD is identical for the non-interacting and interacting limits.}
    \label{fig:1orb_table}
\end{figure*}

\subsection{Molecular Orbital Levels}

In the case of a single orbital, we have just one type of hopping, as in Eq.~\ref{eqn:ordinary_hopping}. The molecular orbitals for different clusters are shown in Fig.~\cref{fig:1orb_table}. Since each site has just a single orbital, an $N$-site cluster has $N$ two-fold degenerate levels.  
For clusters with a well-defined $C_N$-fold rotational symmetry, such as the dimer, triangular and square clusters, molecular orbital basis operators for an $N$-site cluster can be easily defined as 
\begin{equation}
    b^{\dagger}_{l\sigma} = \frac{1}{\sqrt{N}}\sum_{i}c^{\dagger}_{i\sigma}e^{i (2\pi l/N)x_i},
    \label{eqn:mol_basis}
\end{equation}
where $l\in [1,N]$ denotes the quantum number corresponding to rotations of the cluster along the $N$-fold axis of symmetry, and hence, in this case, denotes the different molecular orbitals, and $x_i\in [1,N]$ is a site-index \cite{spin0spin1}. In this basis, the hopping Hamiltonian can be trivially diagonalized and becomes
\begin{equation}
    H_\textrm{non-int} = -2t\sum_{l\sigma}\cos(2 \pi l/N)b^{\dagger}_{l\sigma}b^\pdg_{l\sigma}.
\end{equation}

\subsection{Interaction Hamiltonian}

In the single orbital per site case, $U'$ and $J$ terms in the Hubbard-Kanamori Hamiltonian vanish, leaving us with only the Hubbard interaction. As mentioned in the previous section, for clusters with a well-defined $C_N$-fold rotational symmetry, there is a simple expression, Eq.~\ref{eqn:mol_basis}, for the molecular orbital operators. This means we can express the density-density term associated with $U$ in the molecular orbital basis as
\begin{equation}
    n_{i\uparrow}n_{i\downarrow} = \frac{1}{N}\sum_{lpq}b^{\dagger}_{l\uparrow}b^\pdg_{p\downarrow}b^{\dagger}_{q\downarrow}b^\pdg_{(l+q-p)\downarrow},
    \label{eqn:hint_mol}
\end{equation}
where $l,p,q$ label different molecular orbitals. Now, we can define a molecular orbital spin operator as
\begin{equation}
    S^\alpha_{\textrm{mo},l} = \frac{1}{2}\sum_{\sigma\sigma'}b^\dagger_{l\sigma}\tau^\alpha_{\sigma\sigma'}b_{l\sigma'}
   \label{eqn:sb}
\end{equation}

Along with $N_\textrm{tot} = \sum_{l\sigma}n_{l\sigma}$, this finally gives us an expression for the single-orbital interaction Hamiltonian in the molecular orbital basis as \cite{spin0spin1}
\begin{equation}
\begin{split}
   H_{\textrm{int}} =& -\frac{U}{N}\vec{S}_{\textrm{mo}}^2+\frac{U}{4N}N_\textrm{tot}^2+\frac{U}{2N}N_\textrm{tot}-\frac{U}{N}\sum_{l}n_{l\uparrow}n_{l\downarrow}\\
   & + \frac{U}{N}\sum_{l\neq p}b^{\dagger}_{l\downarrow}b^{\dagger}_{l\uparrow}b_{p\uparrow}b_{(2l-p)\downarrow}\\
& + \frac{U}{N}\sum_{l\neq p \neq q}b^{\dagger}_{q\downarrow}b^{\dagger}_{l\uparrow}b_{p\uparrow}b_{(l+q-p)\downarrow}
\end{split}
\label{eqn:hint_1orb}
\end{equation}
We note that parallels can be drawn between Eq.~\ref{eqn:hint_1orb} and the Hubbard-Kanamori Hamiltonian. Firstly, we see that the $N_\textrm{tot}^2$ term is positive and the $\vec{S}_\textrm{mo}^2$ term is negative, giving rise to similar selection rules as  Eq.~\ref{eqn:hk_gen}. In addition, of the two four-operator terms, the first is a ``pair-clumping" term if $2l \neq p$. This mechanism is shown in Fig.~\cref{fig:1orb_mech}(a), where the $(2l-p)^{th}$ orbital is indicated by $q$. If $2l = p$, then this term becomes a pair-hopping term, as shown in Fig.~\cref{fig:1orb_mech}(b). Similarly, the second of these terms, if $l+q\neq p$, makes two electrons of opposite spin and in different orbitals hop simultaneously to two different empty orbitals respectively. This is shown in Fig.~\cref{fig:1orb_mech}(c). If $l+q =p$, then this becomes a ``pair-spreading" term, which is the mechanism opposite of pair-clumping. This is shown in Fig.~\cref{fig:1orb_mech}(d).

\subsection{Results} 

The results for the single-orbital case are shown in Fig.~\cref{fig:1orb_table}, where the ground state degeneracy (GSD) is listed for each combination of the choice of cluster and electron filling $n_f$. For each combination, the GSD is the same for all finite values of $U$ (the non-interacting GSDs are shown in square brackets when they differ from the interacting case).  

\section{Case-II: Two orbitals Per site} \label{2orb}
\begin{figure}[t]
    \centering
    \includegraphics[width=\columnwidth]{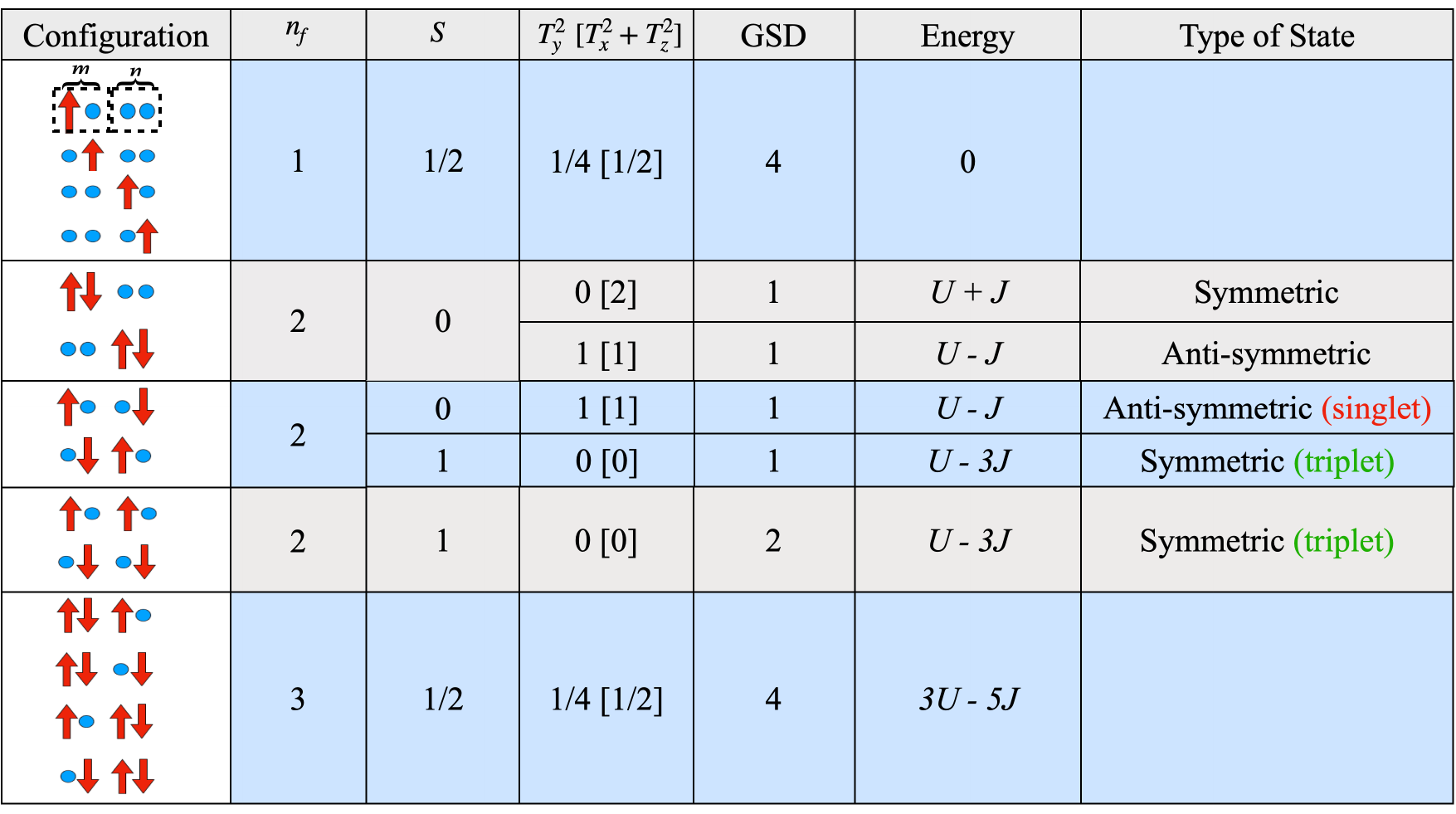}
    \caption{Summary of the two-orbital per site interaction Hamiltonian given in Eq.~\ref{eqn:hint_2orbs_sites} for a single site.}
    \label{fig:2orbs_table}
\end{figure}
\subsection{Molecular Orbital Levels} \label{2orb_nonint}

In the two-orbital case, the non-interacting Hamiltonian $H_{\textrm{non-int}}$ of Eq.~\ref{eqn:multiorb_hopping_gen} is given by:
\begin{equation}
    H_{\textrm{non-int}} = -\sum_{\langle i,j\rangle, \sigma} \boldsymbol{c^\dagger_{i\sigma}}\begin{pmatrix}
t_m & t_{mn} \\
t_{mn} & t_m
\end{pmatrix} \boldsymbol{c_{j\sigma}}
\end{equation}
where $\boldsymbol{c^\dagger_{i\sigma}} = (c^\dagger_{im\sigma}, c^\dagger_{in\sigma})$.  As noted earlier, the off-diagonal hopping $t_{mn}$ breaks $SU(2)$ orbital symmetry, with the orbital character of the bands labeled only by the irreducible representations of $C_2$, $A$ or $B$ (symmetric or anti-symmetric under exchange of the two orbitals). As these are both singly-degenerate, there is no possibility of a non-trivial localized degree of freedom protected solely by orbital symmetry in this two-orbital case. 

\subsection{Interaction Hamiltonian} \label{2orb_int}

We saw in Section \ref{model_methods_int} that the interaction Hamiltonian for the two-orbital case is given by
\begin{equation}
   H_{\textrm{int}}  = \frac{(U-3J)}{2}\sum_i N_i^2 - 2J\sum_i [\vec{S}_i^2+ (T_{i}^y)^2] +\frac{(7J-U)}{2}n_f
   \label{eqn:hint_2orbs_sites}.
\end{equation}
A prominent feature to note here is that only $(T_i^y)^2$ appears in the Hamiltonian, thus orbital isospin $\vec{T}_i^2$ is not conserved. The spectrum of the Hamiltonian for a single site is given in Fig.~\ref{fig:2orbs_table}.

\subsection{Some Select Phase Diagrams} \label{2orb_phasediag}

\subsubsection{Dimer, \texorpdfstring{\(n_f=3\)}{I}}

\begin{figure}
    \centering
    \includegraphics[width=\columnwidth]{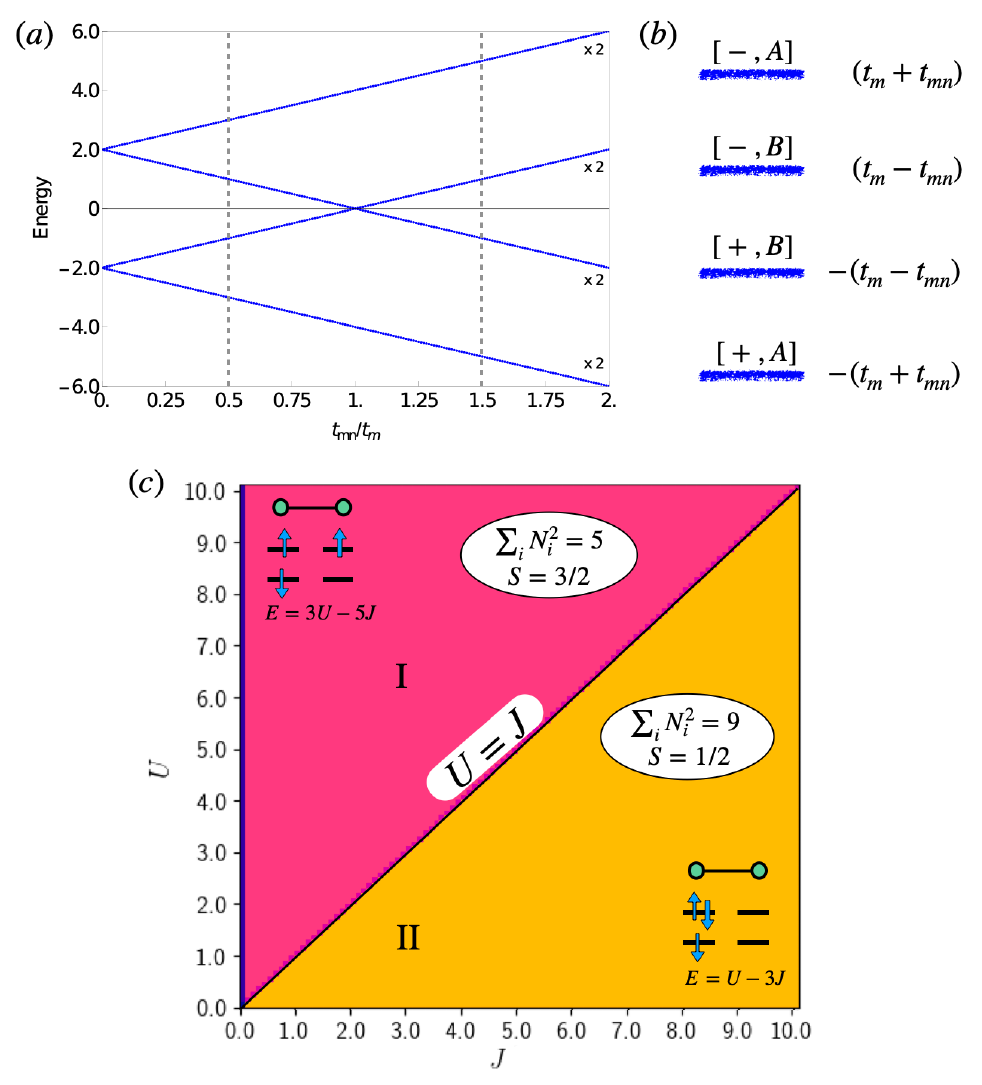}
    \caption{(a) Non-interacting molecular orbital levels for a dimer cluster with two orbitals per site. (b) Single-particle levels with energies and labels indicated: In the label $[G_C,G_O]$, $G_C$ and $G_O$ indicate the irreducible representations of the cluster's spatial and orbital symmetries respectively. (c) $U-J$ phase diagram of $H_\textrm{int}$ only for $n_f = 3$, i.e.~in the absence of hopping.}
    \label{fig:dimer_2orbs_limits}
\end{figure}

Fig.~\cref{fig:dimer_2orbs_limits}(a) shows the non-interacting molecular orbital levels of a dimer cluster with two orbitals per site. The presence of both inter- and intra-orbital hopping gives rise to two regimes: $t_{mn}/t_m<1$ and $t_{mn}/t_m>1$. The dimer cluster has a spatial $i$ point group symmetry and its orbitals have a $C_2$ symmetry, which, as mentioned earlier, we denote as $[i,C_2]$. We consider here the $n_f = 3$ sector. In the non-interacting limit, filling the single-particle levels with 3 electrons gives rise to a two-fold GSD with an effective $S=1/2$ degree of freedom.

\begin{figure}[h]
    \includegraphics[width=\columnwidth]{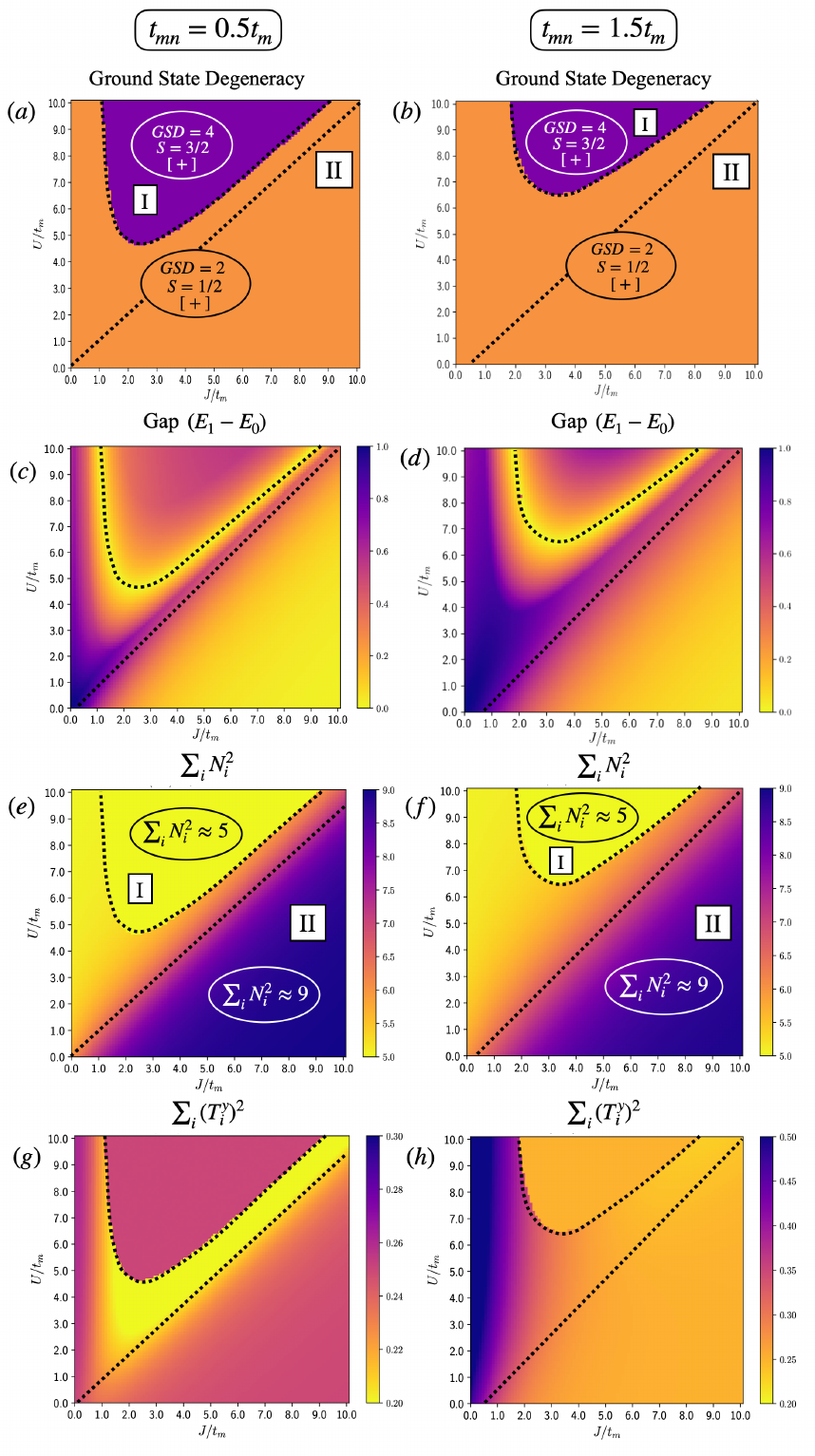}
    \caption{The $U-J$ phase diagrams for a dimer cluster with two orbitals per site and $n_f = 3$. The first column shows the (a) ground state degeneracies (c) gap (e) $\sum_i N_i^2$ and (g) $\sum_i (T^y_i)^2$ plots for $(t_m,t_{mn})=(1.0,0.5)$. The second column shows the (b) ground state degeneracies (d) gap (f) $\sum_i N_i^2$ and (h) $\sum_i (T^y_i)^2$ plots for $(t_m,t_{mn})=(1.0,1.5)$. The GSD plots indicate the GSD, effective spin degree of freedom, and, in square brackets, the inversion quantum number. The dotted boundaries shown in all plots are obtained from the peaks in the second derivative of the ground state energy with respect to $U$ and $J$.}
     \label{fig:dimer_2orbs_panel}
\end{figure}

On the other hand, for $n_f=3$ in the pure interaction limit, there are two possible ways electrons can be distributed among two sites: two electrons on one site and one on the other site, that is, a $(2+1)$ configuration, or, three electrons on one site and none on the other site, that is, a $(3+0)$ configuration. From Eq.~\ref{eqn:hint_2orbs_sites}, we see that a $(2+1)$ configuration is favored in the large-$U$ limit, since this minimizes the $\sum_{i}N_i^2$ term with a value $2^2+1^2=5$, and the $(3+0)$ configuration is favored in the large-$J$ limit with a value $3^2+0^2=9$. This is shown in the pure-interaction phase diagram in Fig.~\cref{fig:dimer_2orbs_limits}(c). Consider the configuration of $(2+1)$ electrons when $U>J$. From Fig.~\cref{fig:2orbs_table}, we see that the energetically favored combination is the presence of an $S=1$ triplet on one site and an $S=1/2$ on the second site. The result of this is an effective $S=3/2$ degree of freedom in region I in Fig.~\cref{fig:dimer_2orbs_limits}(c). Similarly, when $U<J$, $(3+0)$ constitutes the ground state, with an $S=1/2$ on one site and an $S=0$ on the other, resulting in a two-fold ground state degeneracy in region II in Fig.~\cref{fig:dimer_2orbs_limits}(c). Keep in mind that the physically relevant regime is always $U>J$. 

Fig.~\cref{fig:dimer_2orbs_panel} shows phase diagrams for $n_f=3$ in the presence of both interactions and hopping, with phase boundaries indicated. The choices of hoppings $t_{mn}=0.5$ and $t_{mn}=1.5$ are based on the two hopping regimes shown in Fig.~\cref{fig:dimer_2orbs_limits}(a). We see remnants of the pure interaction limit even when hopping is switched on:  the two configurations of electrons being favored in different parameter regimes is seen in Fig.~\cref{fig:dimer_2orbs_panel}(e), but the areas encompassed by $\sum_{i}N_i^2$ values derived from the pure interaction limit have changed due to an interplay of interactions and hopping. This plot is used as a reference to label different regions in the GSD plots in Fig.~\cref{fig:dimer_2orbs_panel}: for example, the purple region in Fig.~\cref{fig:dimer_2orbs_panel}(a) gets label `I' because $\sum_{i}N_i^2 \approx 5$ corresponding to that region in Fig.~\cref{fig:dimer_2orbs_panel}(e); this is same value as that of region I in the pure interaction plot of Fig.~\cref{fig:dimer_2orbs_limits}(c). Similarly, the orange lower-triangular area gets label `II' because $\sum_{i}N_i^2 \approx 9$ in Fig.~\cref{fig:dimer_2orbs_panel}(e) for that area, and this is the same value as that of region II in Fig.~\cref{fig:dimer_2orbs_limits}(c). This labelling convention shall be used in GSD plots for all clusters discussed in the rest of the examples. In addition, the $U=J$ phase boundary remains as is, and the effective spin-degrees of freedom in regions I and II in Fig.~\cref{fig:dimer_2orbs_limits}(a) also follow from the respective regions in Fig.~\cref{fig:dimer_2orbs_limits}(c). 

As we increase hopping, we see that the region I shrinks, and the two-fold GSD occupies a larger area. In addition, the $U=J$ line shifts away from the origin (Fig.~\cref{fig:dimer_2orbs_panel}(b)). The non-interacting limit (that is, the origin in the plot: $U=0,J=0$) is one which favors a more delocalized distribution of electrons, hence smoothly connecting to the region where a $(2+1)$ configuration forms the ground state. It also favors a lower total spin, since the lowest energy levels get filled sequentially, as opposed to a higher effective spin degree of freedom favored in the pure interaction limit (due to Hund's rules). As a result, there is a competition between favoring a higher effective spin and a lower effective spin when $U>J$. As hopping is increased, there is a tendency of the system to approach the behavior of the non-interacting limit, hence shrinking the region with higher effective spin. 

\subsubsection{Trimer, \texorpdfstring{\(n_f=5\)}{I}}

\begin{figure}[h]
    \centering
    \includegraphics[width=\columnwidth]{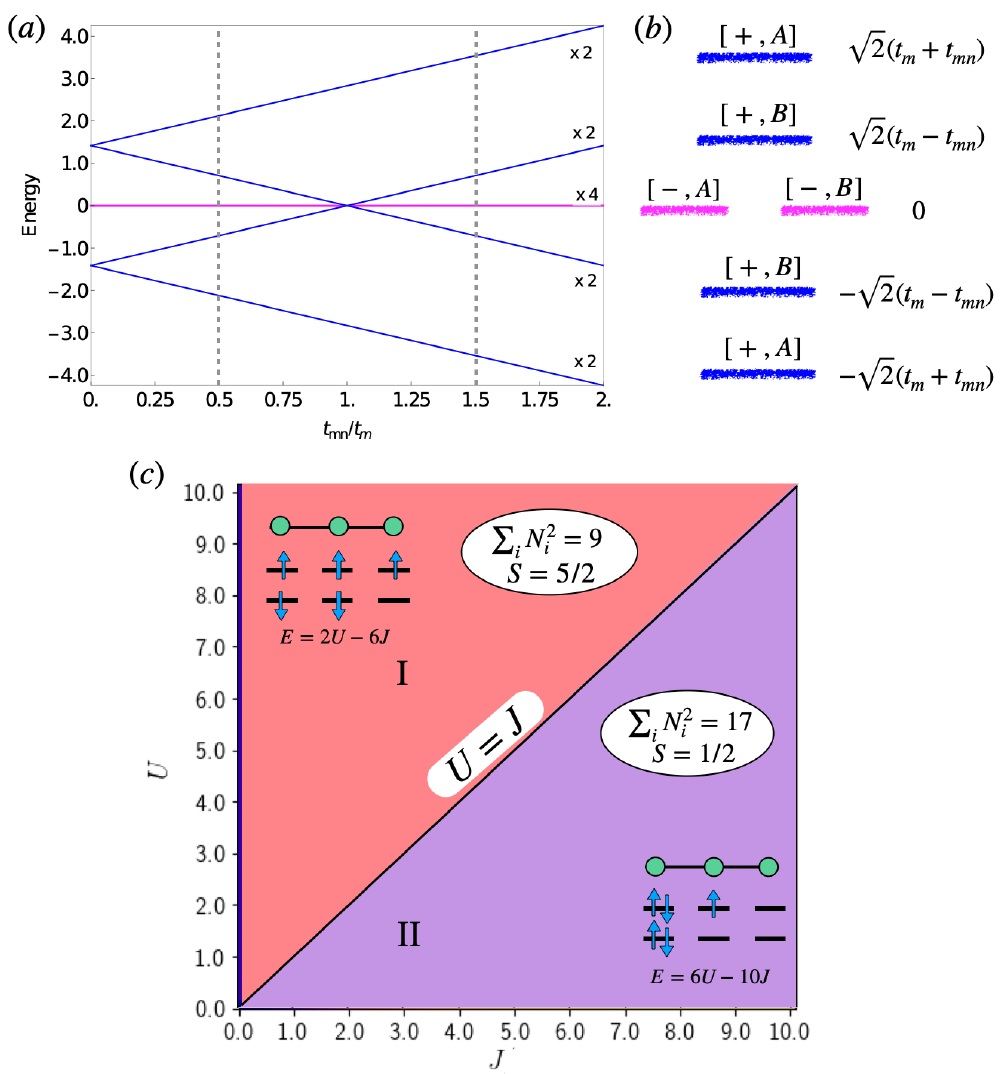}
    \caption{(a) Non-interacting molecular orbital levels for a trimer cluster with two orbitals per site. (b) Single-particle levels with energies indicated, ordered assuming $t_{mn} < t_m$. (c) $U-J$ phase diagram of $H_\textrm{int}$ only for $n_f = 5$, i.e.~in the absence of hopping.}
    \label{fig:trimer_2orbs_limits}
\end{figure}

Fig.~\cref{fig:trimer_2orbs_limits}(a) shows the non-interacting molecular orbital levels of a trimer cluster with two orbitals per site. As with the dimer cluster, there are two regimes: $t_{mn}/t_m<1$ and $t_{mn}/t_m>1$. The trimer cluster has a $[i,C_2]$ symmetry. A distinct feature of the trimer molecular orbital levels are the zero-energy $[-,A]$ and $[-,B]$ levels. These levels are protected by inversion symmetry of the trimer cluster. In this section, we have chosen to show the $n_f = 5$ sector as an example. In the non-interacting limit, we see that filling the single-particle levels with five electrons gives rise to a four-fold ground state degeneracy, with an effective $S=1/2$ degree of freedom.

In the pure interaction limit, there are many possible ways in which five electrons can be distributed across three sites of the trimer. Among these configurations, we see from Eq.~\ref{eqn:hint_2orbs_sites} that a $(2+2+1)$ configuration is favored in the large-$U$ limit (since this minimizes the $\sum_{i}N_i^2$ term) and the $(4+1+0)$ configuration is favored in the large-$J$ limit. This is shown in the pure-interaction plot of Fig.~\cref{fig:trimer_2orbs_limits}(c).

\begin{figure}
    \hspace{-0.43cm}
    \includegraphics[width=8.7cm,height=13cm]{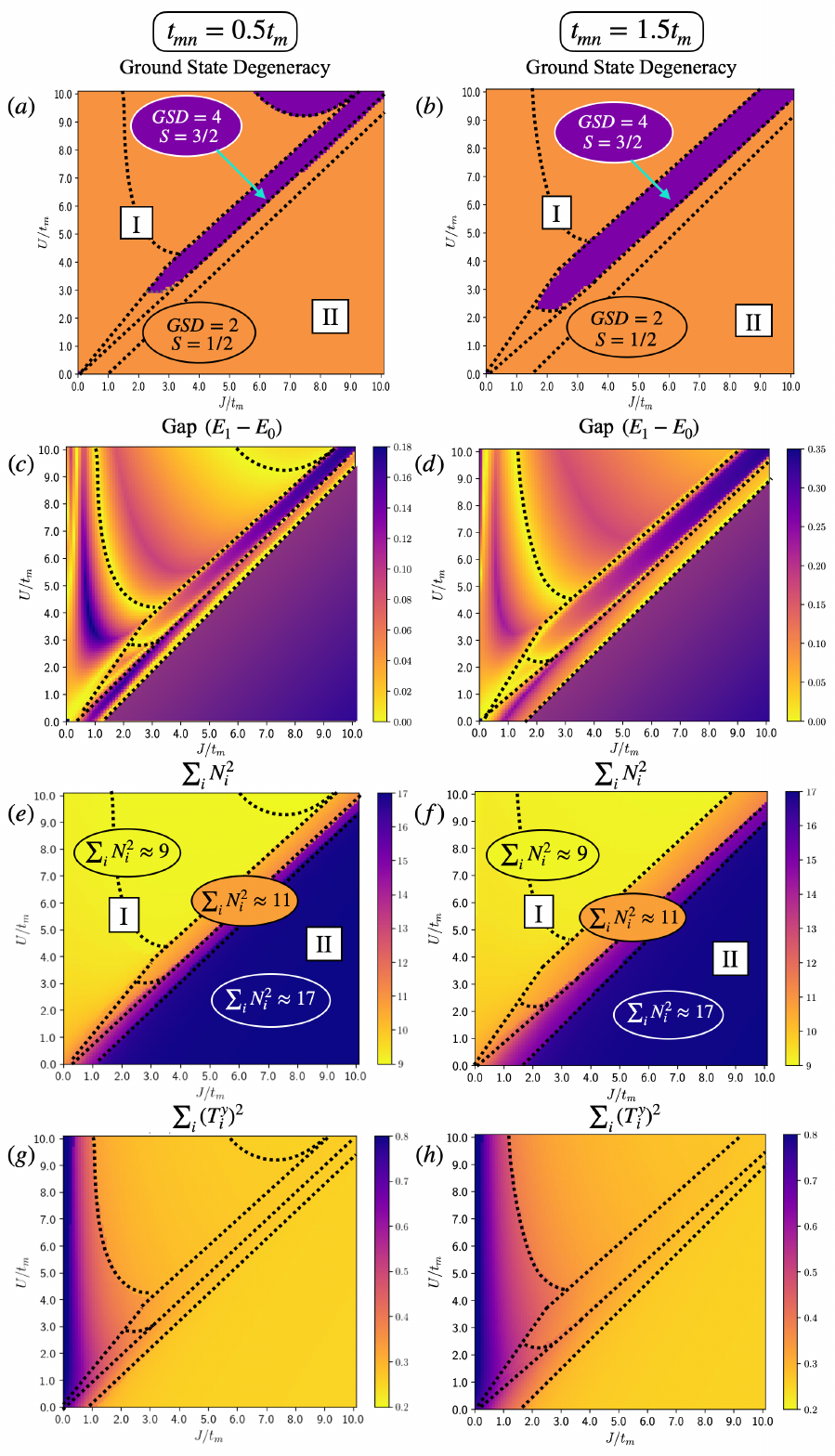}
     \caption{The $U-J$ phase diagrams for a trimer cluster with two orbitals per site and $n_f = 5$. The first column shows the (a) ground state degeneracies (c) gap (e) $\sum_i N_i^2$ and (g) $\sum_i (T^y_i)^2$ plots for $(t_m,t_{mn})=(1.0,0.5)$. The second column shows the (b) ground state degeneracies (d) gap (f) $\sum_i N_i^2$ and (h) $\sum_i (T^y_i)^2$ plots for $(t_m,t_{mn})=(1.0,1.5)$.}
    \label{fig:trimer_2orbs_panel}
\end{figure}

Fig.~\cref{fig:trimer_2orbs_panel} shows the phase diagrams for $n_f=5$ in the presence of both interactions and hopping ($t_{mn}=0.5$ and $t_{mn}=1.5$), with phase boundaries indicated.
In Fig.~\cref{fig:trimer_2orbs_panel}(a), there is a plethora of phases in different regions. Firstly, we note remnants from the pure interaction limit: in addition to the $U=J$ phase boundary, the electronic configurations being favored in different parameter regimes can be seen in Fig.~\cref{fig:trimer_2orbs_panel}(e), although the areas corresponding to regions I and II from the pure interaction limit have now shrunk due to an interplay of interactions and hopping. Secondly,  as hopping is introduced, we see a new region with $S=3/2$ opening up around the $U=J$ line, corresponding to the $(3+1+1)$ configuration (purple region in Fig.~\cref{fig:tetra_2orbs_panel}(a)).

As we increase hopping, we see that the purple region slightly expands. In addition, we can observe the non-interacting limit (that is, the origin in the plot: $U=0,J=0$) is now surrounded by areas where $S=1/2$. This reiterates the fact that, as hopping is increased, there is a tendency of the system to approach the behavior of the non-interacting limit, and that the intermediate regime between pure hopping and pure interaction limits is one with multiple phases.

\subsubsection{Triangle, \texorpdfstring{\(n_f=7\)}{I}}

\begin{figure}
    \centering
    \includegraphics[width=7.5cm,height=8.0cm]{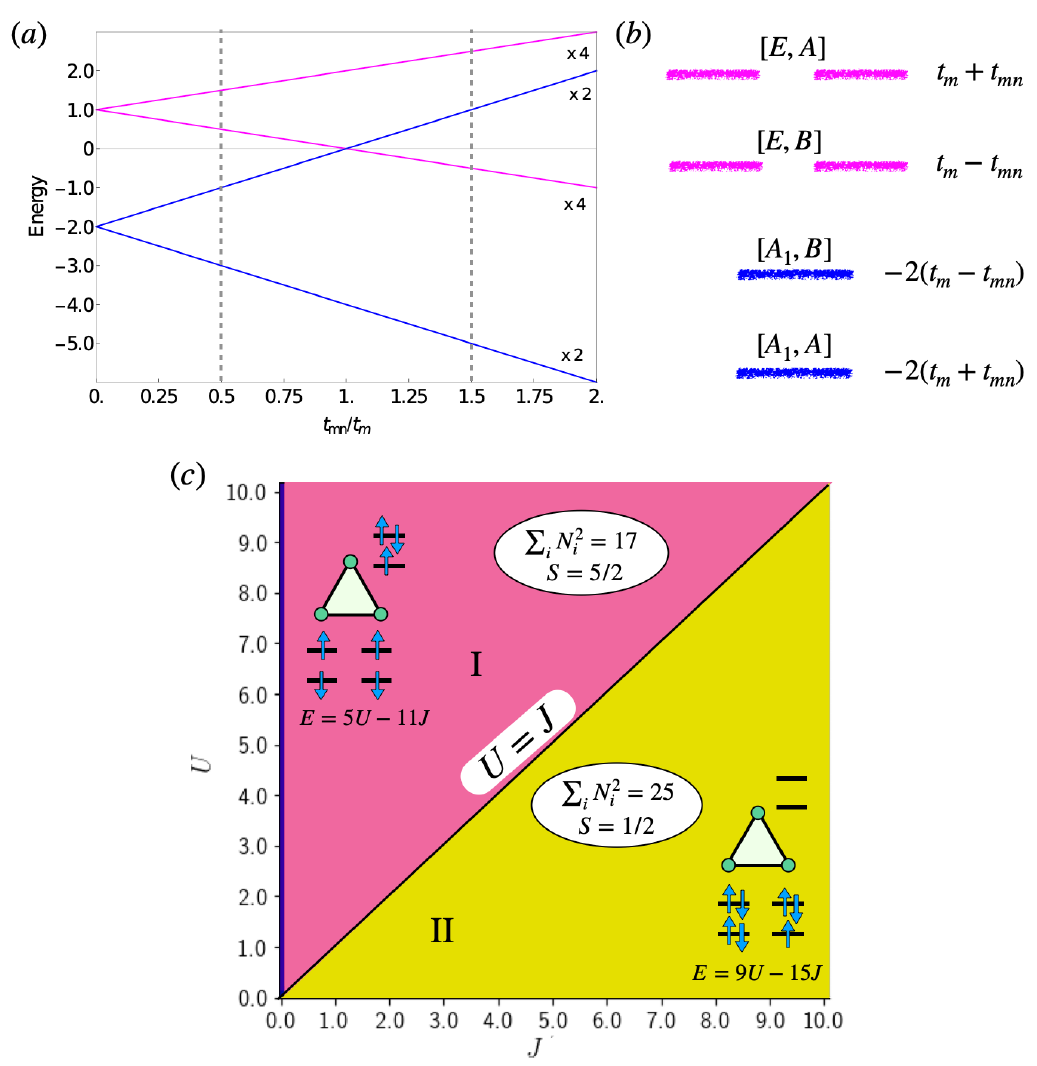}
      \caption{(a) Non-interacting molecular orbital levels for a triangular cluster with two orbitals per site. (b) Single-particle levels with energies indicated, ordered assuming $t_{mn} < t_m$. (c) $U-J$ phase diagram of $H_\textrm{int}$ only for $n_f = 7$, i.e.~in the absence of hopping.}
    \label{fig:triangle_2orbs_limits}
\end{figure}

Fig.~\cref{fig:triangle_2orbs_limits}(a) shows the non-interacting molecular orbital levels of a triangle cluster with two orbitals per site. We have, as before, the regime of $t_{mn}/t_m<1$ and $t_{mn}/t_m>1$. The triangular cluster has a $[C_{3v},C_2]$ symmetry. A distinct feature of the triangle molecular orbital levels are the two-fold degenerate $[E,A]$ and $[E,B]$ levels. These levels are protected by the $C_3$ symmetry of the triangular cluster. We consider the $n_f = 7$ sector here as an example. In the non-interacting limit, filling the single-particle levels with 7 electrons gives rise to a four-fold ground state degeneracy when $t_{mn}/t_m<1$ and a two-fold ground state degeneracy when $t_{mn}/t_m>1$. 

There are many different ways of arranging seven electrons among three sites; a $(3+2+2)$ configuration is favored in the large-$U$ limit (since this minimizes the $\sum_{i}N_i^2$ term) and a $(4+3+0)$ configuration is favored in the large-$J$ limit, as shown in the pure-interaction plot of Fig.~\cref{fig:triangle_2orbs_limits}(c). Consider first the region I. From Fig.~\cref{fig:2orbs_table}, we see that the energetically favored combination is the presence of an $S=1$ triplet on the two sites with two electrons, and an $S=1/2$ on the third site (in accordance with Hund's rules). The result of this is an overall effective $S=5/2$ degree of freedom. Similarly, in region II, an $S=1/2$ on one site and an $S=0$ on the other two results in an overall $S=1/2$ degree of freedom.

\begin{figure}
    \hspace{-0.43cm}
    \includegraphics[width=9.0cm,height=13cm]{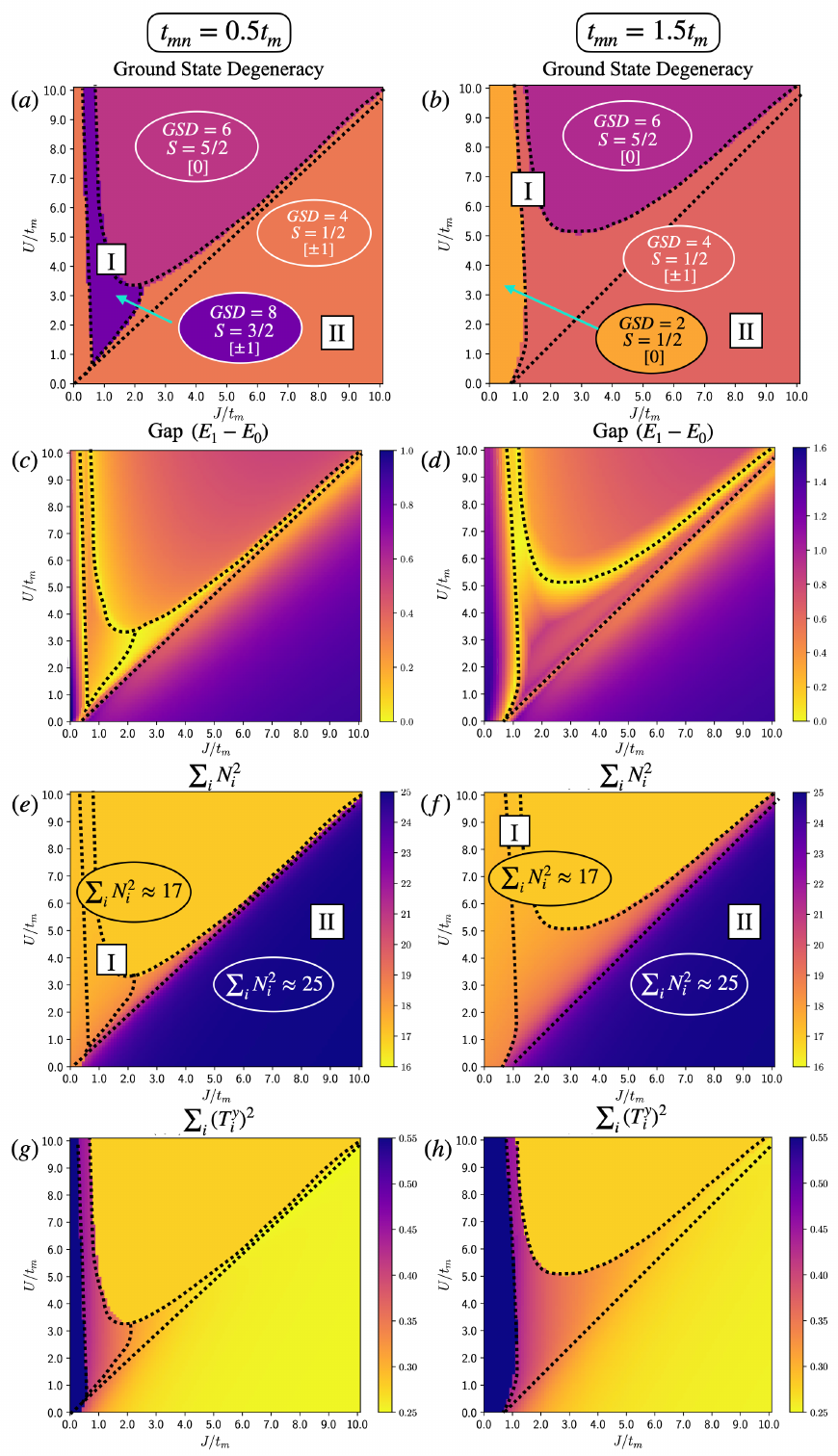}
     \caption{The $U-J$ phase diagrams for a triangular cluster with two orbitals per site and $n_f = 7$. The first column shows the (a) ground state degeneracies (c) gap (e) $\sum_i N_i^2$ and (g) $\sum_i (T^y_i)^2$ plots for $(t_m,t_{mn})=(1.0,0.5)$. The second column shows the (b) ground state degeneracies (d) gap (f) $\sum_i N_i^2$ and (h) $\sum_i (T^y_i)^2$ plots for $(t_m,t_{mn})=(1.0,1.5)$. The quantum number indicated in square brackets corresponds to rotation about the cluster's $C_3$ axis.}
     \label{fig:triangle_2orbs_panel}
\end{figure}

Fig. \cref{fig:triangle_2orbs_panel} shows the phase diagrams for $n_f=7$ with both interactions and hopping. In Fig. \cref{fig:triangle_2orbs_panel}(a), 
we see a variety of phases in region I and II.  Different configurations of electrons being favored in different parameter regimes is seen in the $\sum_{i}N_i^2$ plot in Fig.~\cref{fig:triangle_2orbs_panel}(e), although the areas encompassed by values close to those of the pure interaction limit have changed due to an interplay of interactions and hopping. Another remnant of the pure interaction limit is the $U=J$ phase boundary; in addition, the effective spin degrees of freedom in region I (pink area) and region II also follow from their respective pure interaction counterparts in Fig.~\cref{fig:triangle_2orbs_limits}(c). Note that while the degeneracy is purely due to the spin degree of freedom in the pink area of region I, the GSD in other areas and regions arise due to a combination of spin and spatial symmetries. 

As we increase hopping to $t_{mn}/t_m>1$, we see that region I shrinks, and region II expands (Fig.~\cref{fig:triangle_2orbs_panel}(b)). In addition, the $U=J$ line shifts away from the origin, and the non-interacting limit (that is, the origin in the plot: $U=0,J=0$) smoothly connects to the new area with an $S=1/2$ degree of freedom, in tune with the preferred ground state in the non-interacting limit. We hence observe a tendency of the system trying to approach this limit, in the lower left area of  Fig.~\cref{fig:triangle_2orbs_panel}(b). 

\subsubsection{Tetramer, \texorpdfstring{\(n_f=7\)}{I}}

\begin{figure}
    \centering
    \includegraphics[width=7.5cm,height=8.0cm]{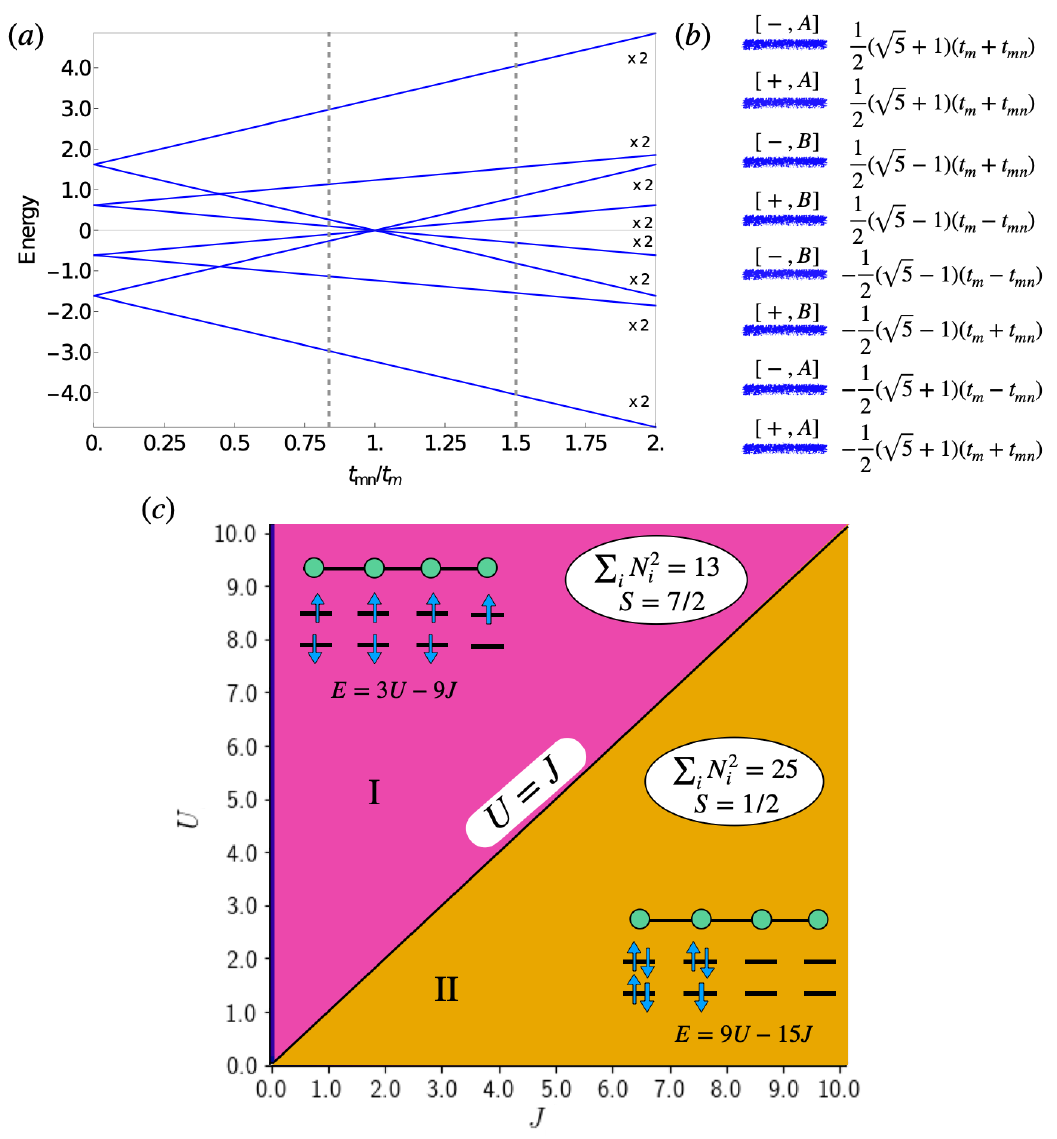}
       \caption{(a) Non-interacting molecular orbital levels for a tetramer cluster with two orbitals per site. (b) Single-particle levels with energies indicated, ordered assuming $t_{mn} < t_m$. (c) $U-J$ phase diagram of $H_\textrm{int}$ only for $n_f = 7$, i.e.~in the absence of hopping.}
    \label{fig:tetramer_2orbs_limits}
\end{figure}

Fig.~\cref{fig:tetramer_2orbs_limits}(a) shows the non-interacting
molecular orbital levels of a tetramer cluster with two orbitals per site. The presence of both inter- and intra-orbital hopping gives rise to three regimes: $t_{mn}<1/\sqrt{5}$, $1/\sqrt{5}<t_{mn}<1$, and $t_{mn}/t_m>1$. Of these, only the second and third regimes are highlighted (for reasons discussed below). The tetramer cluster has a $[i,C_2]$ symmetry. We show here $n_f = 7$ as an example. In the non-interacting limit, filling the single-particle levels with 7 electrons gives rise to a doubly degenerate ground state with an effective $S=1/2$ degree of freedom.

In the absence of hopping, among all the different ways that 7 electrons can be distributed among four sites, the $(2+2+2+1)$ configuration is favored in the large-$U$ limit and the $(4+3+0+0)$ configuration is favored in the large-$J$ limit, as shown in the pure-interaction plot of Fig.~\cref{fig:tetramer_2orbs_limits}(c).

\begin{figure}
    \hspace{-0.43cm}
    \includegraphics[width=9.0cm,height=14cm]{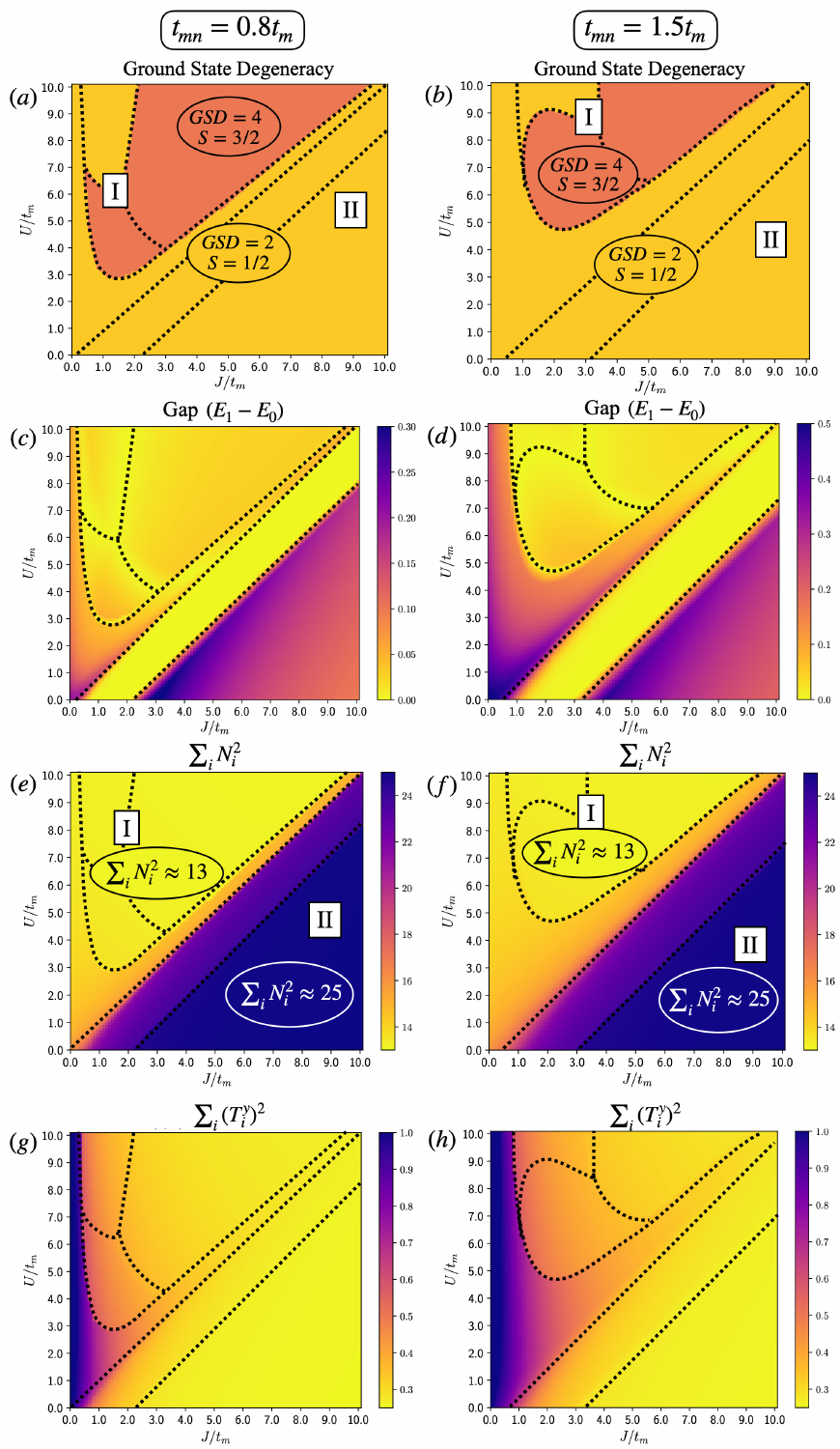}
      \caption{The $U-J$ phase diagrams for a tetramer cluster with two orbitals per site and $n_f = 7$. The first column shows the (a) ground state degeneracies (c) gap (e) $\sum_i N_i^2$ and (g) $\sum_i (T^y_i)^2$ plots for $(t_m,t_{mn})=(1.0,0.5)$. The second column shows the (b) ground state degeneracies (d) gap (f) $\sum_i N_i^2$ and (h) $\sum_i (T^y_i)^2$ plots for $(t_m,t_{mn})=(1.0,1.5)$.}
    \label{fig:tetra_2orbs_panel}
\end{figure}

We now switch on $U$ and $J$, and study how these ground states evolve. In the first regime of $t_{mn}<1/\sqrt{5}$, there is a uniform two-fold degeneracy arising from an effective $S=1/2$ everywhere. Hence, plots for this regime are not shown. Fig.~\cref{fig:tetra_2orbs_panel} shows the phase diagrams for $n_f=7$, with phase boundaries indicated. The choices of hoppings $t_{mn}=0.8$ and $t_{mn}=1.5$ are based on the two hopping regimes of Fig.~\cref{fig:tetramer_2orbs_limits}(a).

In Fig.~\cref{fig:tetra_2orbs_panel}(a), there are a variety of phases in regions I and II, either with a two-fold or a four-fold GSD. The two regions of different electronic configurations are confirmed by Fig.~\cref{fig:tetra_2orbs_panel}(e), although the areas encompassed by a ``pure" $(2+2+2+1)$ configuration (region I) and a ``pure" $(4+3+0+0)$ configuration (region II) have changed. 

When $t_{mn}$ is increased, we see that region I shrinks, and that the area of two-fold degeneracy (region II) expands.  In addition, the $U=J$ line shifts away from the origin (Fig. \cref{fig:tetra_2orbs_panel}(b)), with the non-interacting limit now smoothly connected to the $GSD=2$ region. 

\subsubsection{Tetrahedron, \texorpdfstring{\(n_f=6\)}{I}}

\begin{figure}
    \centering\includegraphics[width=7.5cm,height=8.0cm]{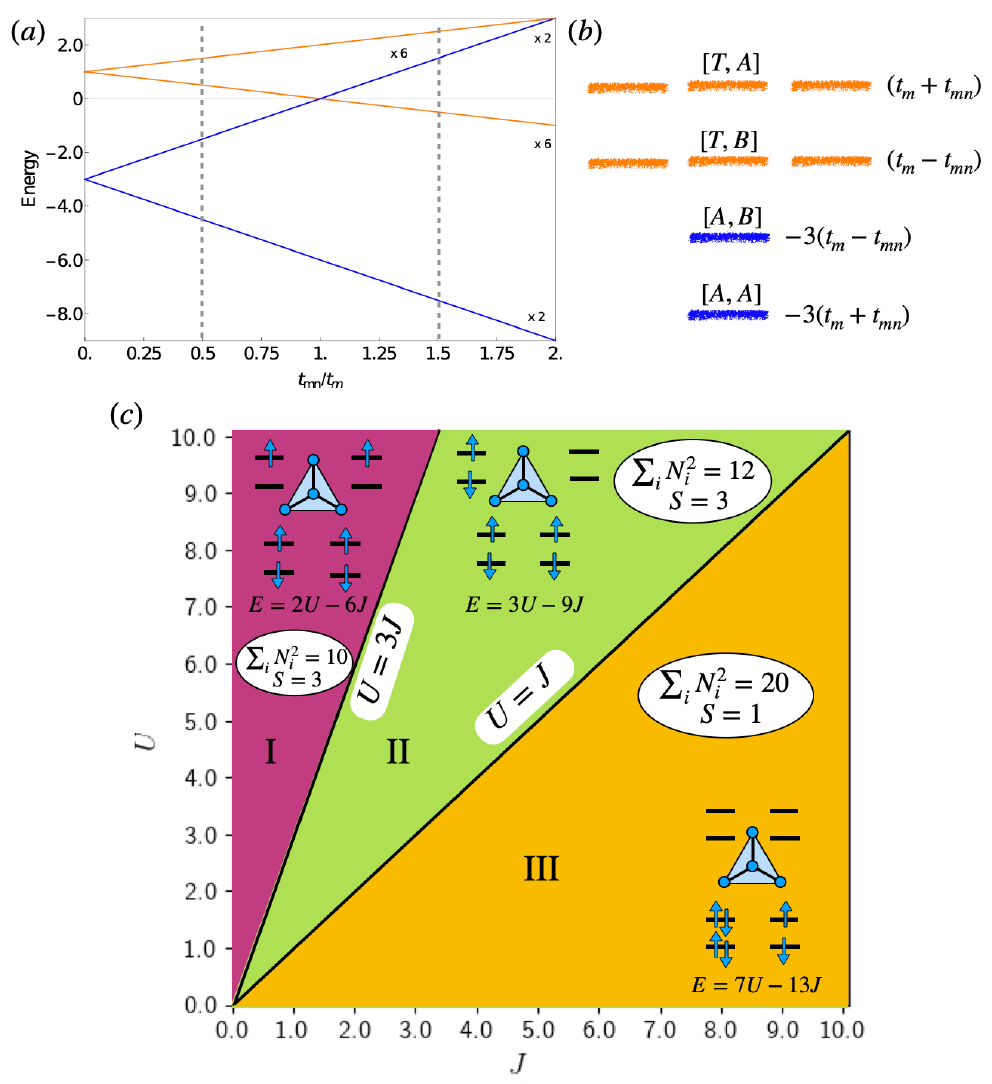}
       \caption{(a) Non-interacting molecular orbital levels for a tetrahedral cluster with two orbitals per site. (b) Single-particle levels with energies indicated, ordered assuming $t_{mn} < t_m$. (c) $U-J$ phase diagram of $H_\textrm{int}$ only for $n_f = 6$, i.e.~in the absence of hopping.}
    \label{fig:tetrahedron_2orbs_limits}
\end{figure}

Fig.~\cref{fig:tetrahedron_2orbs_limits}(a) shows the non-interacting molecular orbital levels of a tetrahedral cluster with two orbitals per site. We have, as before, the regime of $t_{mn}/t_m<1$ and $t_{mn}/t_m>1$. The tetrahedral cluster has a $[T_d,C_2]$ symmetry. A distinct feature of these molecular orbital levels are the $[T,A]$ and $[T,B]$ levels with a three-fold degeneracy each. We show here $n_f = 6$ as an example. In the non-interacting limit, filling the single-particle levels with 6 electrons gives rise to a fifteen-fold ground state degeneracy, with individual states that can have either $S=0$ or $S=1$ effective degrees of freedom.

In the opposite limit of pure interactions, there are many ways of arranging 6 electrons on a tetrahedon: the $(2+2+1+1)$ configuration is favored in the large-$U$ limit and the $(4+2+0+0)$ configuration is favored in the large-$J$ limit. The $(2+2+2+0)$ configuration is favored in the intermediate regime. This is shown in the pure-interaction plot of Fig.~\cref{fig:tetrahedron_2orbs_limits}(c).

\begin{figure}
    \hspace{-0.43cm}
    \includegraphics[width=9cm,height=14cm]{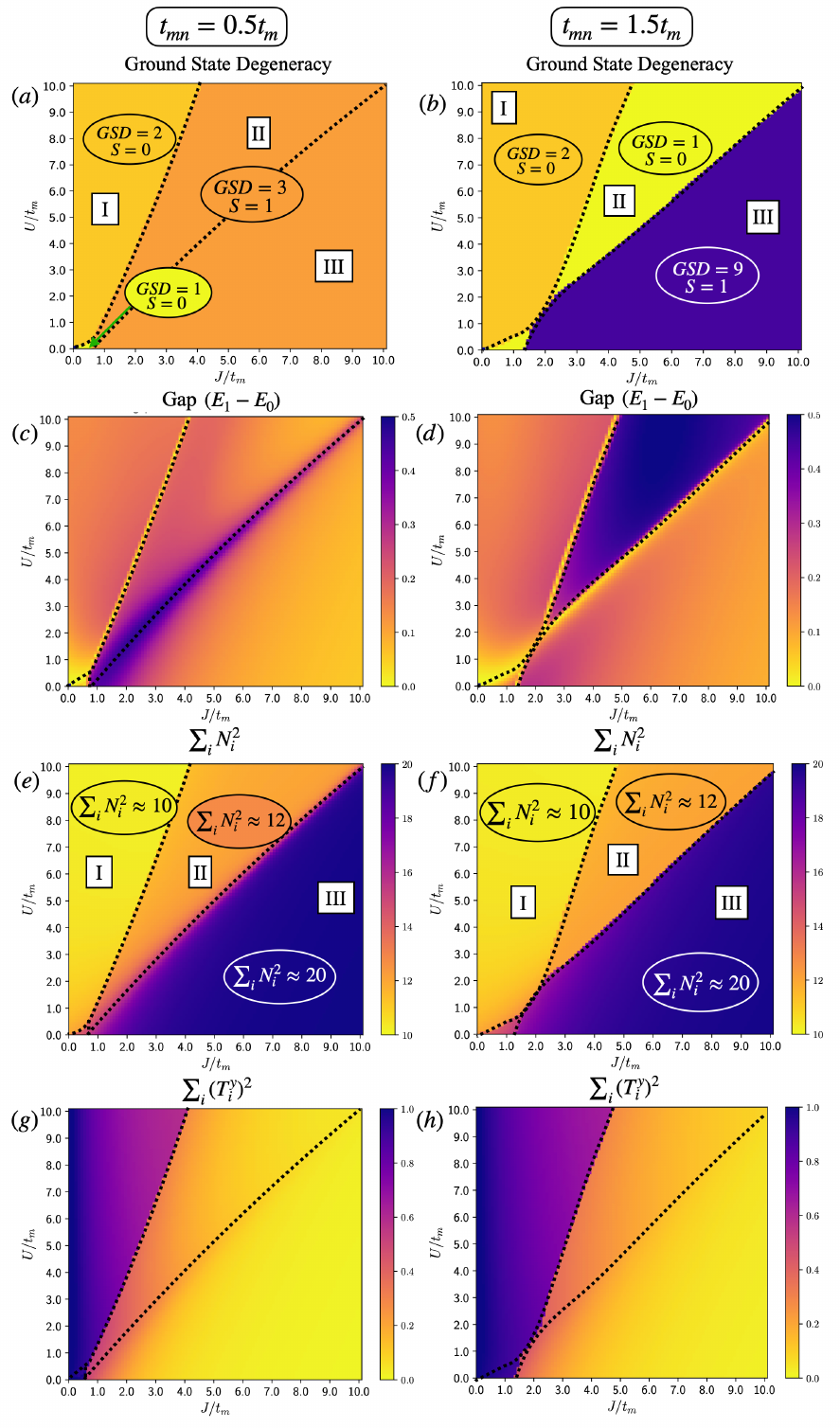}
      \caption{The $U-J$ phase diagrams for a tetrahedral cluster with two orbitals per site and $n_f = 6$. The first column shows the (a) ground state degeneracies (c) gap (e) $\sum_i N_i^2$ and (g) $\sum_i (T^y_i)^2$ plots for $(t_m,t_{mn})=(1.0,0.5)$. The second column shows the (b) ground state degeneracies (d) gap (f) $\sum_i N_i^2$ and (h) $\sum_i (T^y_i)^2$ plots for $(t_m,t_{mn})=(1.0,1.5)$. The quantum number indicated in square brackets corresponds to rotation about the clusters $C_3$ axis.}
    \label{fig:hedron_2orbs_panel}
\end{figure}

Fig. \cref{fig:hedron_2orbs_panel} shows the phase diagrams for $n_f=6$ in the intermediate regime of both interactions and hopping. Note that while the degeneracy is purely due to spin degrees of freedom in region II and region III in Fig.~\cref{fig:hedron_2orbs_panel}(a), the ground state in region I is a spin singlet and its two-fold GSD is instead due to spatial symmetry of the cluster. 

As hopping increases to $t_{mn}/t_m>1$, region II expands and now has a single unique ground state (Fig. \cref{fig:hedron_2orbs_panel}(b)). In addition to the cluster's $T_d$ symmetry protecting the two-fold GSD in region I, we also see a spatial contribution to the GSD in region III which, when combined with the $S=1$ spin contribution, results in an overall nine-fold degeneracy. We also observe that although the non-interacting point is not smoothly connected to any neighboring regions, adding a small $U$ or a small $J$ to this point gives an $S=0$ ground state, and ground states with higher effective spin degrees of freedom can only be realized at larger $J$.

\subsubsection{Square, \texorpdfstring{\(n_f=11\)}{I}}

\begin{figure}
    \centering\includegraphics[width=7.5cm,height=8.0cm]{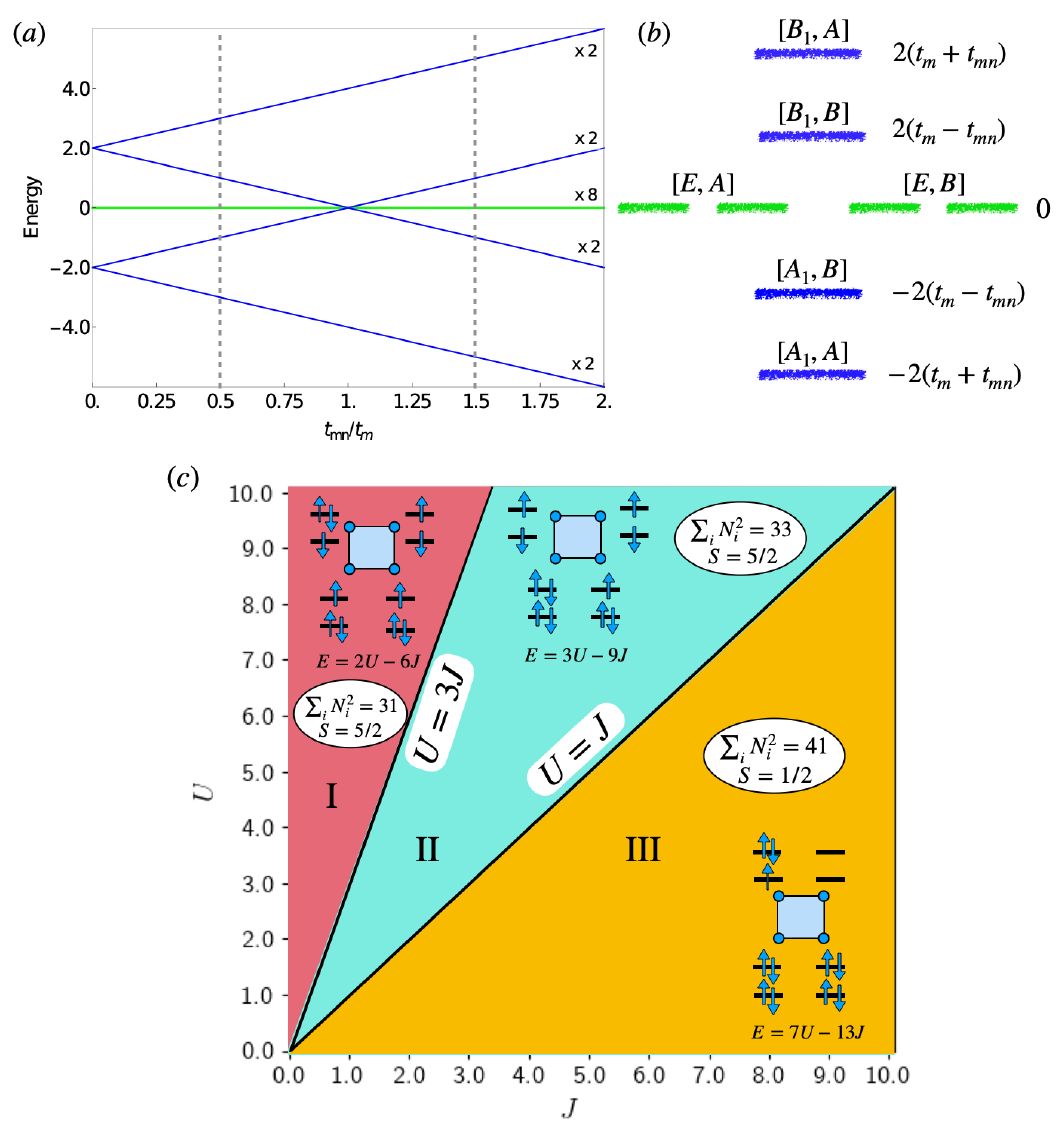}
       \caption{(a) Non-interacting molecular orbital levels for a square cluster with two orbitals per site. (b) Single-particle levels with energies indicated, ordered assuming $t_{mn} < t_m$. (c) $U-J$ phase diagram of $H_\textrm{int}$ only for $n_f = 11$, i.e.~in the absence of hopping.}
    \label{fig:square_2orbs_limits}
\end{figure}

Fig.~\cref{fig:square_2orbs_limits}(a) shows the non-interacting molecular orbital levels of a square cluster with two orbitals per site. Once again, there are two distinct regimes with $t_{mn}/t_m<1$ and $t_{mn}/t_m>1$. The square cluster has a $[C_{4v},C_2]$ symmetry. A distinct feature of the square molecular orbital levels are the zero-energy $[E,A]$ and $[E,B]$ levels. These levels can be split by breaking the cluster $C_{4v}$ down to $C_{2v}$.  
We have chosen to show $n_f = 11$ here as an example. In the non-interacting limit, filling the single-particle levels with 11 electrons gives rise to an eight-fold ground state degeneracy with an effective $S=1/2$ degree of freedom.

In the pure interaction limit, a $(3+3+3+2)$ configuration is favored in the large-$U$ limit and a $(4+4+3+0)$ configuration is favored in the large-$J$ limit. The intermediate region has a $(4+3+2+2)$ configuration as its ground state, as shown in the pure-interaction plot of Fig.~\cref{fig:square_2orbs_limits}(c). 
 
\begin{figure}
    \hspace{-0.43cm}
    \includegraphics[width=9cm,height=14cm]{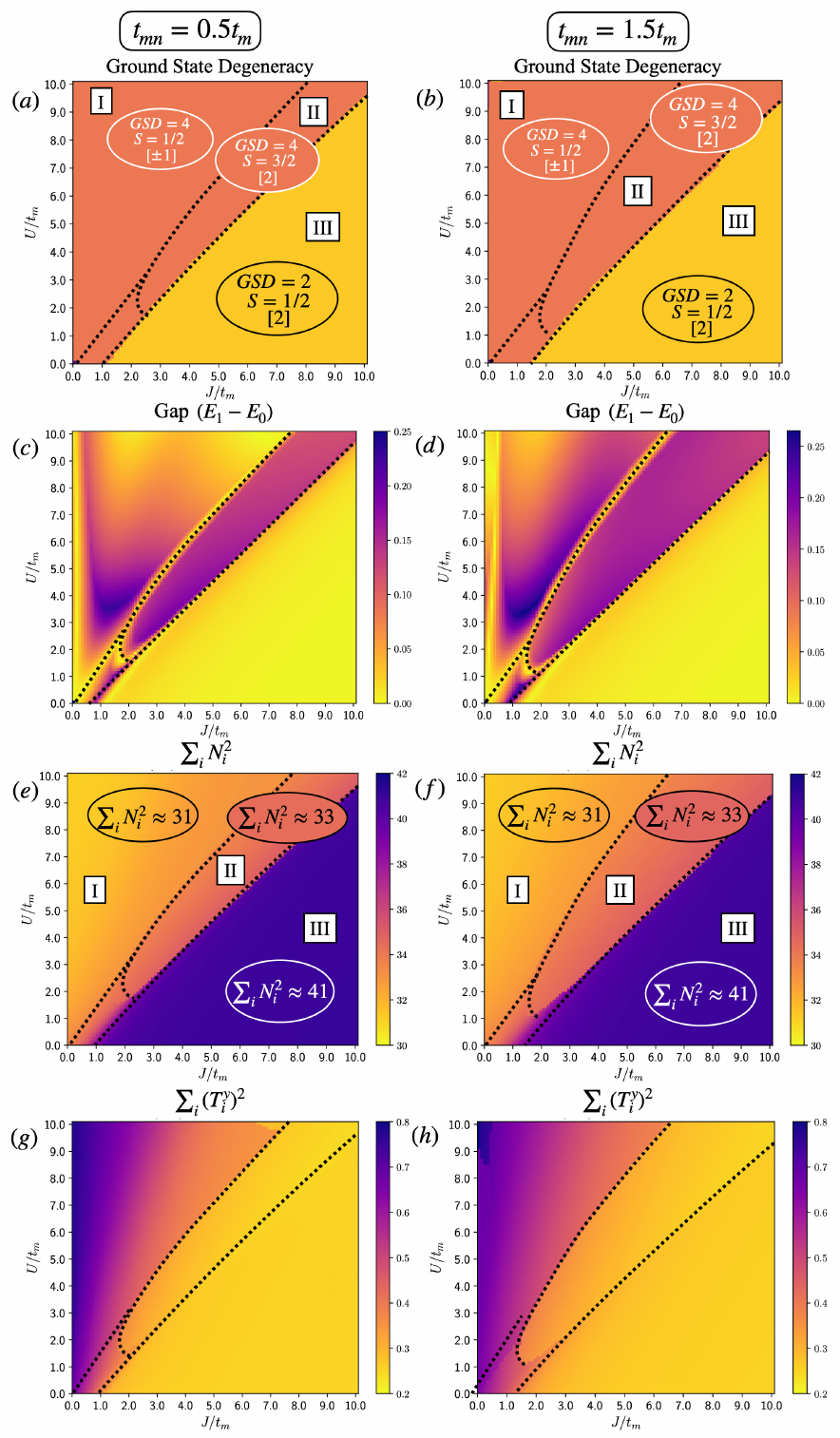}
      \caption{The $U-J$ phase diagrams for a square cluster with two orbitals per site and $n_f = 11$. The first column shows the (a) ground state degeneracies (c) gap (e) $\sum_i N_i^2$ and (g) $\sum_i (T^y_i)^2$ plots for $(t_m,t_{mn})=(1.0,0.5)$. The second column shows the (b) ground state degeneracies (d) gap (f) $\sum_i N_i^2$ and (h) $\sum_i (T^y_i)^2$ plots for $(t_m,t_{mn})=(1.0,1.5)$. The quantum number indicated in square brackets corresponds to rotation about the cluster's $C_4$ axis.}
    \label{fig:square_2orbs_panel}
\end{figure}

Fig.~\cref{fig:square_2orbs_panel} shows the phase diagrams for $n_f=11$, with both interactions and hoppings switched on. As with the other cases, we see remnants from the pure interaction limit even though hoppings are now introduced: that is, regimes where different electronic configurations constitute the ground state, as seen in Fig.~\cref{fig:square_2orbs_panel}(e), although the areas encompassed have shifted. Note that while the degeneracy is purely due to the spin degree of freedom in region II and region III in Fig.~\cref{fig:square_2orbs_panel}(a), the GSD in region I arises due to a combination of spin symmetry and the $C_4$ rotational symmetry of the cluster.

As we increase hopping, there are a few observations to make: regions I and III gradually shrink, whereas there is a very slight increase in the area of region II. The non-interacting point is distinct from the surrounding regions; however, adding a small $U$ or $J$ leads to a four-fold ground state degeneracy, with different effective spin degrees of freedom.

\section{Case-III: Three Orbitals per site} \label{3orb}

\subsection{Molecular Orbital Levels} \label{3orb_nonint}

In the three-orbital case, the non-interacting Hamiltonian $H_{\textrm{non-int}}$ is given by:
\begin{equation}
    H_{\textrm{non-int}} = -\sum_{\langle i,j\rangle, \sigma} \boldsymbol{c^\dagger_{i\sigma}}\begin{pmatrix}
t_m & t_{mn} & t_{mn}\\
t_{mn} & t_m & t_{mn}\\
t_{mn} & t_{mn} & t_{m}
\end{pmatrix} \boldsymbol{c_{j\sigma}}
\end{equation}
where $\boldsymbol{c^\dagger_{i\sigma}} = (c^\dagger_{im\sigma}, c^\dagger_{in\sigma},c^\dagger_{ip\sigma})$. As already mentioned, the inter-orbital hopping $t_{mn}$ breaks the continuous $SO(3)$ orbital symmetry down to a discrete $C_{3v}$ symmetry. In addition to its singly-degenerate irreducible representations, $A_1$ and $A_2$, $C_{3v}$ also contains a two-fold degenerate irreducible representation, $E$. In stark contrast to the two-orbital case, this thus allows for the possibility of a non-Kramers doublet protected purely by orbital symmetry.  

\subsection{Interaction Hamiltonian} \label{3orb_int}

From Section \ref{model_methods_int} we saw that the Hubbard-Kanamori Hamiltonian for a cluster with three orbitals per site is given by
\begin{equation}
\begin{split}
    H_{\textrm{int}} =& \frac{(U-3J)}{2}\sum_{i}N_i^2-2J\sum_i[ \vec{S}_i^2 + \left(\vec{L}_i/2\right)^2 ]\\ 
    & +\frac{(8J-U)}{2}n_f 
    \label{eqn:hint_3orbs_sites}
\end{split}
\end{equation}
 
\begin{figure}[h]
    \centering
    \includegraphics[width=8.5cm,height=5cm]{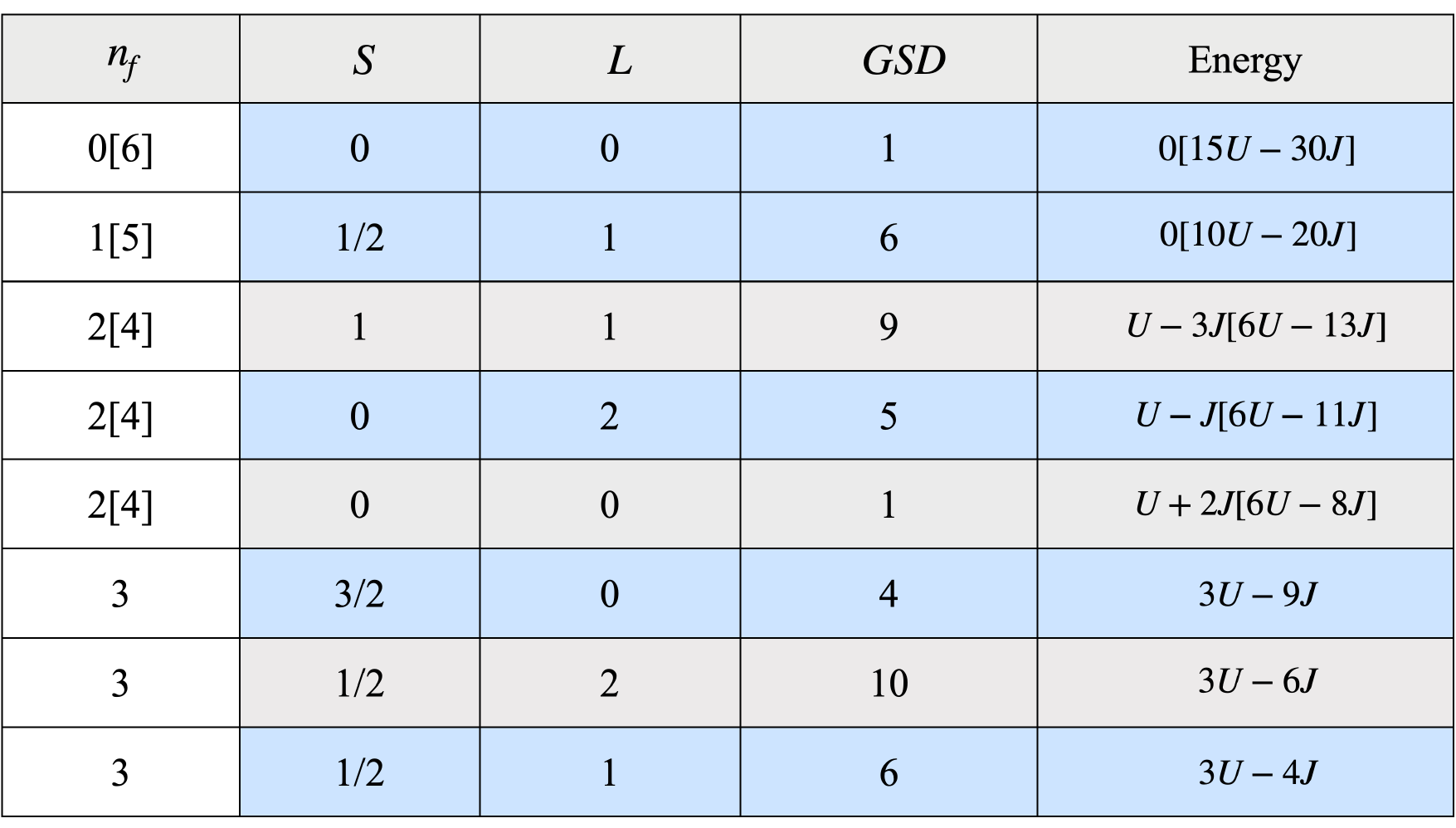}
    \caption{Summary of the three-orbital per site interaction Hamiltonian given in Eq.~\ref{eqn:hint_3orbs_sites}, for a single site. The energies given in square brackets correspond to the $n_f$ given in square brackets.}
    \label{fig:3orbs_table}
\end{figure}

The spectrum of this Hamiltonian for a single site is shown in Fig.~\cref{fig:3orbs_table} \cite{Georges_2013}. Note here that the angular momentum at each site, $\vec{L}_i^2$, is conserved, in contrast with the two-orbital case.

\subsection{Some select Phase Diagrams} \label{3orb_phasediag}

\subsubsection{Dimer, \texorpdfstring{\(n_f=8\)}{I}}

\begin{figure}
    \centering
    \includegraphics[width=7.5cm,height=7.5cm]{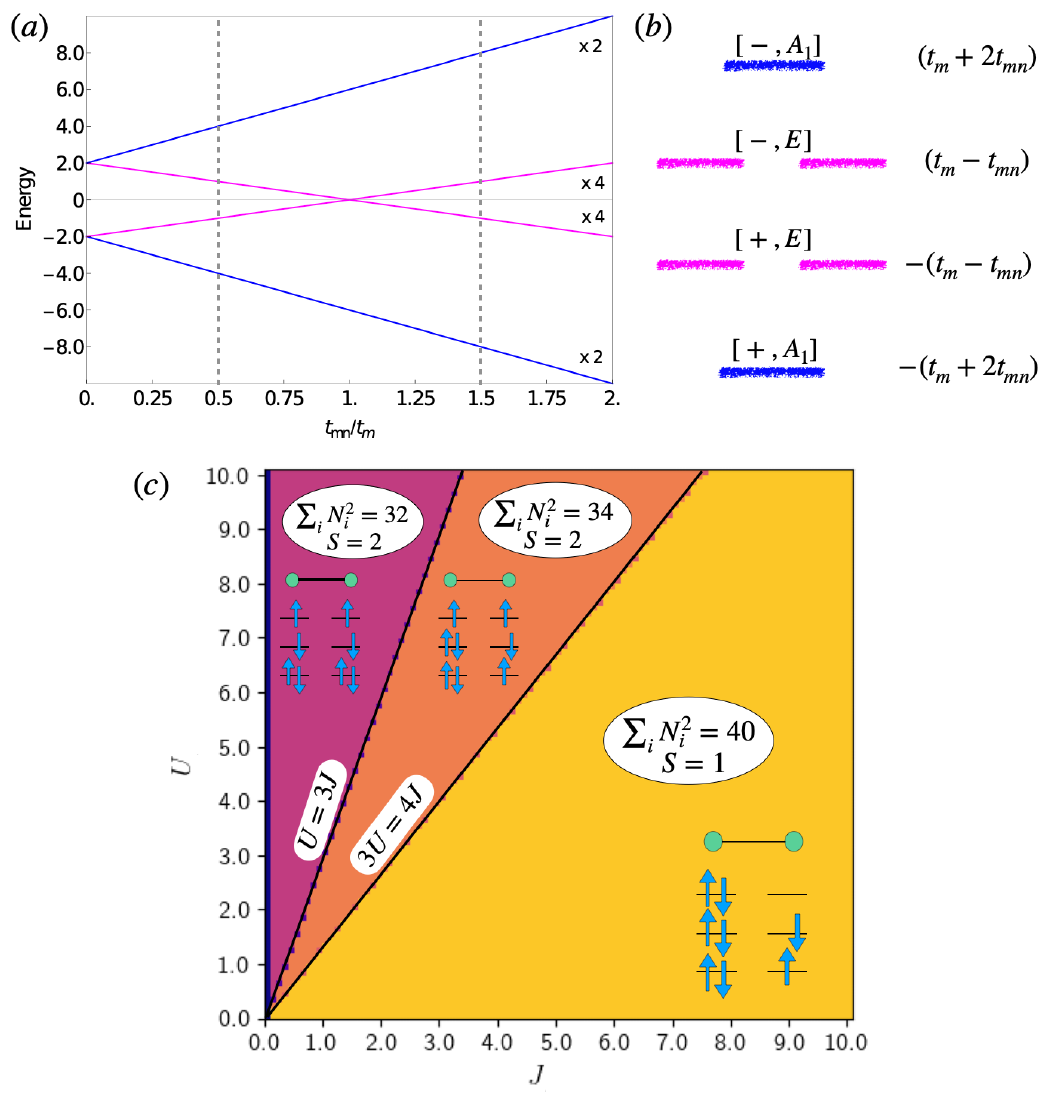}
       \caption{(a) Non-interacting molecular orbital levels for a dimer cluster with three orbitals per site. (b) Single-particle levels with energies indicated, ordered assuming $t_{mn} < t_m$. (c) $U-J$ phase diagram of $H_\textrm{int}$ only for $n_f = 8$, i.e.~in the absence of hopping.}
    \label{fig:dimer_3orbs_limits}
\end{figure}

\begin{figure}
    \hspace{-0.43cm}
    \includegraphics[width=9cm,height=13cm]{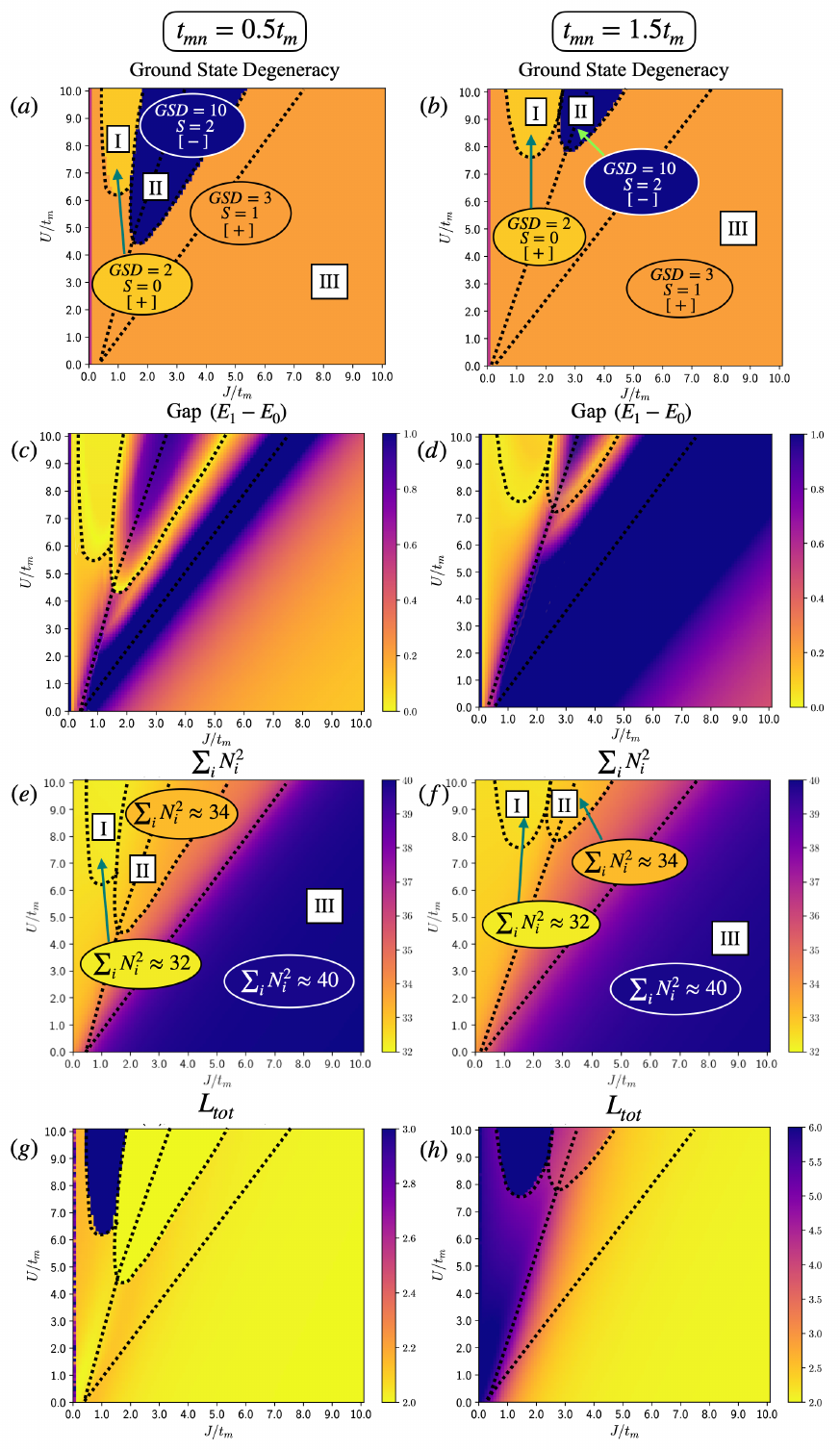}
      \caption{The $U-J$ phase diagrams for a dimer cluster with three orbitals per site and $n_f = 8$. The first column shows the (a) ground state degeneracies (c) gap (e) $\sum_i N_i^2$ and (g) $(\sum_i L_i)^2$ plots for $(t_m,t_{mn})=(1.0,0.5)$. The second column shows the (b) ground state degeneracies (d) gap (f) $\sum_i N_i^2$ and (h) $(\sum_i L_i)^2$ plots for $(t_m,t_{mn})=(1.0,1.5)$. Indicated in square brackets in (a) and (b) is the inversion quantum number.}
    \label{fig:dimer_3orbs_panel}
\end{figure}

Fig.~\cref{fig:dimer_3orbs_limits}(a) shows the non-interacting molecular orbital levels of a dimer cluster with three orbitals per site. As always, there is the $t_{mn}/t_m<1$ and $t_{mn}/t_m>1$ regimes. The dimer cluster has a $[i,C_{3v}]$ symmetry. A distinct feature of these molecular orbital levels are the $[\pm,E]$ levels. These levels are protected by the orbital $C_3$ symmetry (Fig.~\cref{fig:hopping_mech}(c)). We show here $n_f = 8$ as an example. In the non-interacting limit, filling the single-particle levels with 8 electrons gives rise to a six-fold ground state degeneracy, with individual states that can either have an $S=0$ or $S=1$ effective degree of freedom.

In the pure interaction limit, for $n_f=8$,  there are only three possible ways the electrons can be distributed among the two sites of the cluster. Of these, it can be shown that the $(4+4)$ configuration is favored in region I, the $(6+2)$ configuration in region II, and the $(5+3)$ configuration in region III (see Fig.~\cref{fig:dimer_3orbs_limits}(c)).

Fig.~\cref{fig:dimer_3orbs_panel} shows the phase diagrams for $n_f=8$, with both interactions and hoppings. 
Note that while the degeneracy is purely due to the spin degree of freedom in region III in Fig.~\cref{fig:dimer_3orbs_panel}(a), the GSD in regions I and II have different origins: the two-fold GSD in region I is protected entirely by the $C_{3v}$ symmetry of the orbitals, whereas the GSD in region II arises due to a combination of both spin and orbital symmetry.

As the hopping increases, regions I and II shrink, and region III expands, and the $U=J$ line shifts away from the origin (see Fig.~\cref{fig:dimer_3orbs_panel}(b)).The non-interacting limit smoothly connects to the region with $GSD=3$. As with the previous cases, we observe that, as hopping is increased, there is a tendency of the system to approach the behavior of the non-interacting limit.

\subsubsection{Trimer, \texorpdfstring{\(n_f=7\)}{I}}

\begin{figure}[t]
\includegraphics[width=7.5cm,height=8cm]{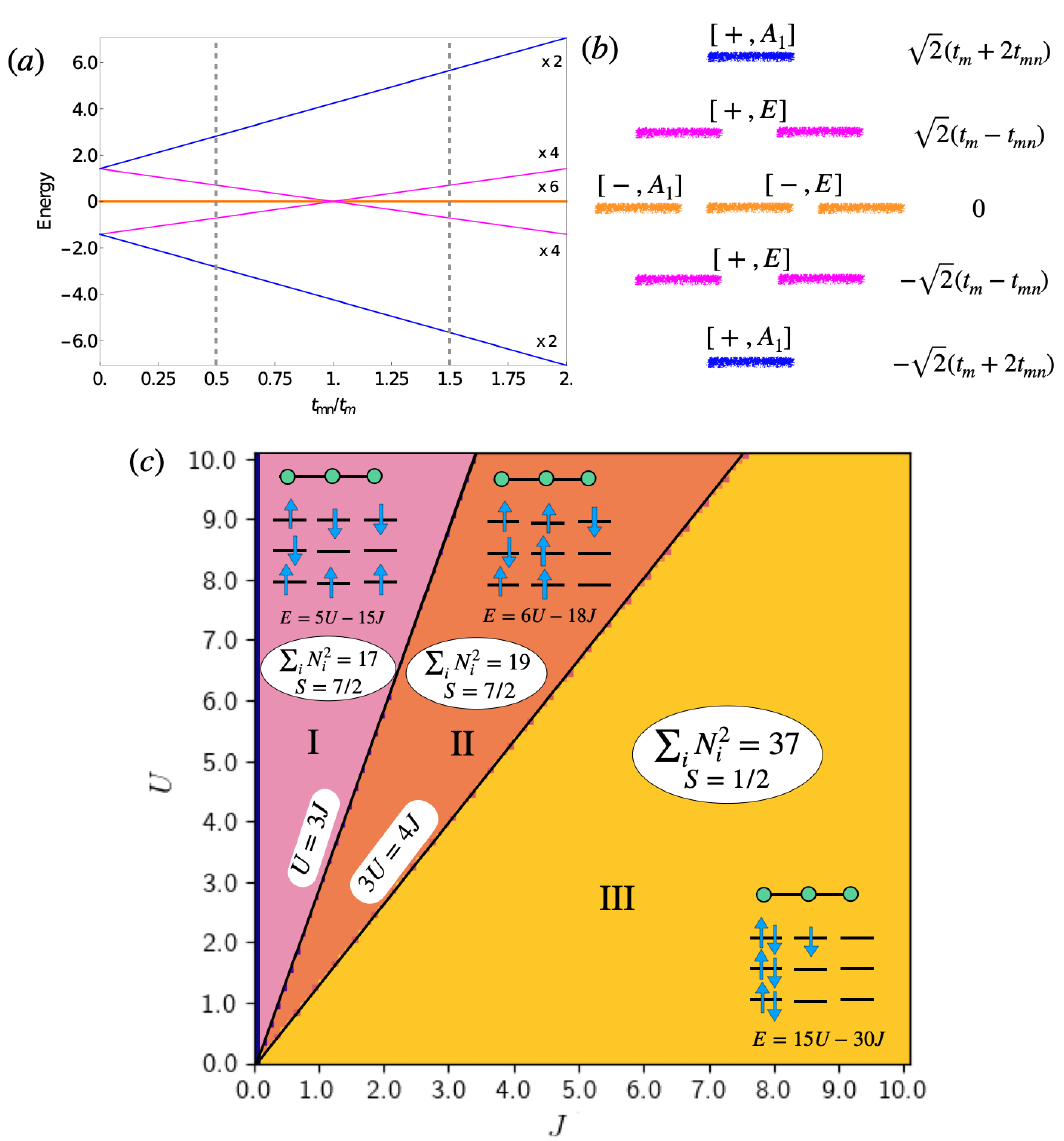}
  \caption{(a) Non-interacting molecular orbital levels for a trimer cluster with three orbitals per site. (b) Single-particle levels with energies indicated, ordered assuming $t_{mn} < t_m$. (c) $U-J$ phase diagram of $H_\textrm{int}$ only for $n_f = 7$, i.e.~in the absence of hopping.}
    \label{fig:trimer_3orbs_limits}
\end{figure}

\begin{figure}[h]
    \hspace*{-0.5cm}
    \includegraphics[width=9.1cm,height=14cm]{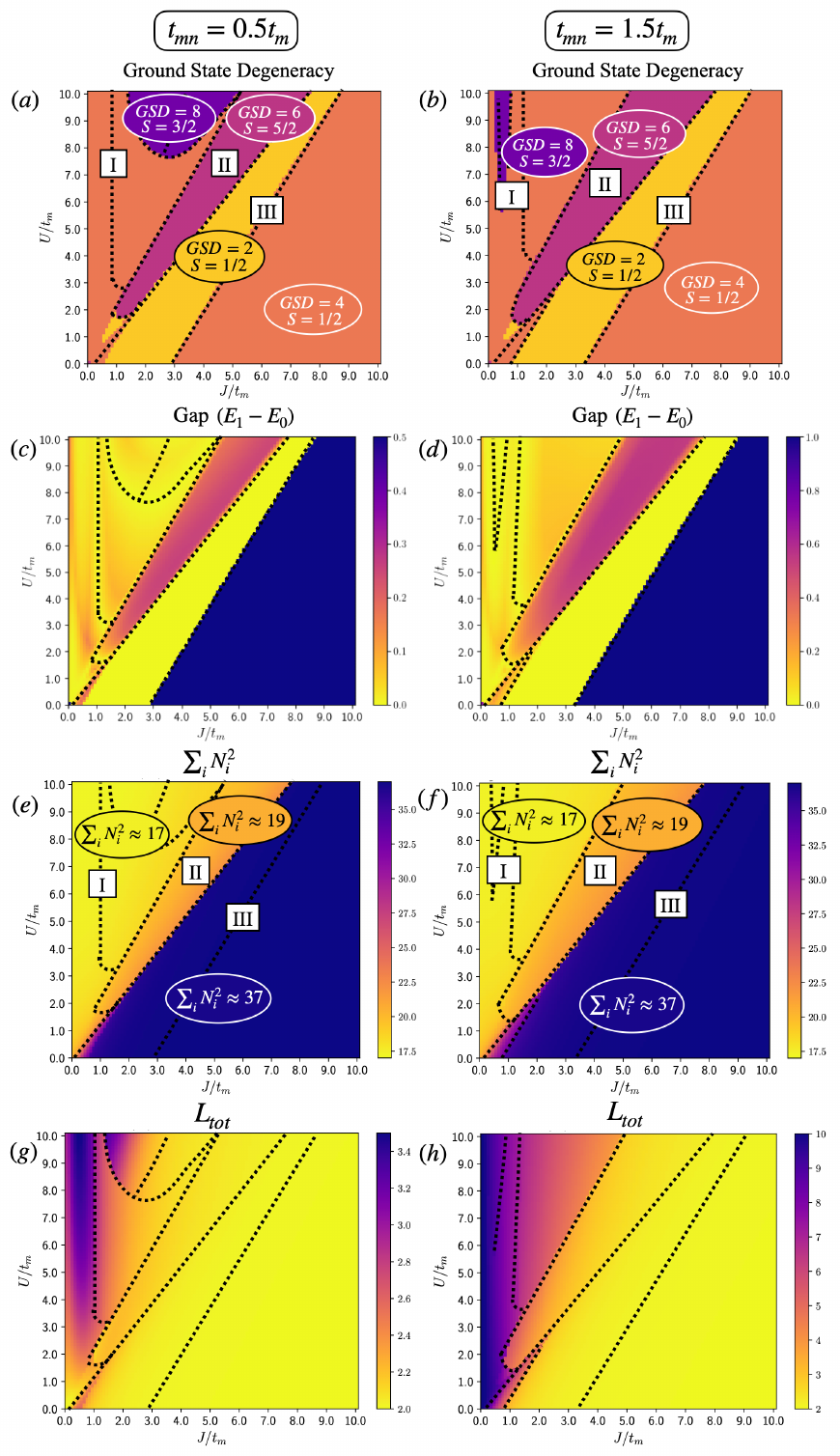}
     \caption{The $U-J$ phase diagrams for a trimer cluster with three orbitals per site and $n_f = 7$. The first column shows the (a) ground state degeneracies (c) gap (e) $\sum_i N_i^2$ and (g) $(\sum_i L_i)^2$ plots for $(t_m,t_{mn})=(1.0,0.5)$. The second column shows the (b) ground state degeneracies (d) gap (f) $\sum_i N_i^2$ and (h) $(\sum_i L_i)^2$ plots for $(t_m,t_{mn})=(1.0,1.5)$.}
    \label{fig:trimer_3orbs_panel}
\end{figure}

Fig.~\cref{fig:trimer_3orbs_limits}(a) shows the non-interacting molecular orbital levels of a trimer cluster with three orbitals per site, with again the distinct $t_{mn}/t_m<1$ and $t_{mn}/t_m>1$ regimes. The trimer cluster has a $[i,C_{3v}]$ symmetry. A distinct feature of the molecular orbital levels are the two-fold degenerate $[+,E]$ levels and the zero-energy $[-,A],[-,E]$ levels. The $[+,E]$ bands are protected by the orbital $C_3$ symmetry. The zero-energy $[-,A]$ and $[-,E]$ levels are protected by inversion symmetry. We have chosen to show $n_f = 7$ as an example. In the non-interacting limit, filling the single-particle levels with 7 electrons gives rise to a six-fold degenerate ground state with an $S=1/2$ degree of freedom.

In the pure interaction limit, for $n_f=7$, the configurations shown in the pure-interaction plot of Fig.~\cref{fig:trimer_3orbs_limits}(c) are favored in the respective parameter regimes. Switching on $U$ and $J$, Fig.~\cref{fig:trimer_3orbs_panel} shows the phase diagrams for $n_f=7$, with phase boundaries indicated. In Fig.~\cref{fig:trimer_3orbs_panel}(a), we see that many new regions have emerged. Moreover, the configurations being favored are confirmed by Fig.~\cref{fig:trimer_3orbs_panel}(e), with the $(3+2+2)$ (region I), $(3+3+1)$ (region II), and $(6+1+0)$ (region III) configurations visible in the values of $\sum_i N_i^2$. Note that the GSD in region I and part of region III arises due to a combination of spin and orbital symmetries of the cluster, whereas GSD elsewhere arises purely due to spin.

As we increase hopping, we see that the region with $S=3/2$ has drastically shrunk, and a larger area of the plot is occupied with different regions having an $S=1/2$ degree of freedom. In addition, the $U=J$ line has shifted very slightly away from the origin.

\section{Discussion and Outlook} \label{discussion}

In this work, we explored a relatively simple set of model cluster Hamiltonians for a wide variety of cluster geometries and electron fillings. As is evident from the phase diagrams, the models display a rich diversity of behavior and complexity. However, there are a number of unifying features. In the purely interacting limit, the phase diagrams usually display two or three distinct regions, distinguished by their value of $\avg{\sum_i N_i^2}$. These can be understood from the form of Eq.~\ref{eqn:hk_gen}, the Hubbard-Kanamori Hamiltonian written in terms of $N_i^2, \vec{S}_i^2$ and $Q_i^2$, and the lowest energy states of single sites for different fillings given in Figs.~\ref{fig:2orbs_table} and \ref{fig:3orbs_table}. The full phase diagrams, including hopping and interactions, are more complex, with for example numerous gap closings occurring throughout each phase diagram. However, it is worth noting that, even here, there are only ever $2$-$4$ distinct ground states, i.e.~with distinct GSD, spin and lattice irreducible representation. These can be understood either as arising from breaking down the high degeneracies found in the corresponding interaction-only phase diagrams, or, in other cases, from filling the non-interacting molecular orbital levels. These unifying features lead us to the three key insights mentioned in the introduction, namely (i) the additional cluster Hund's rule associated with the $\sum_i N_i^2$ term in the Hubbard-Kanamori Hamiltonian (a term which is a simple constant and plays no role in conventional single-site Mott insulators), (ii) the appearance of ground states which are best understood from the viewpoint of the purely interacting Hamiltonian, as opposed to simple filling of molecular orbital levels, and (iii) the relative scarcity of non-trivial degeneracies protected by lattice or orbital symmetries (which has important implications for real materials as we will discuss next). Our relatively simple set of models thus provides for a more overarching understanding of cluster ground state degeneracies, and the interplay between electron hopping and interactions, while maintaining the key ingredients of cluster Mott materials.

We now briefly comment on the likely impact of additional terms required for an accurate quantitative comparison with realistic candidate materials, which we have thus far neglected (note that these additions are all at the level of the non-interacting Hamiltonian, as the interacting Hamiltonian used here already accounts for realistic multi-orbital Hubbard-Kanamori interactions). Crystal field terms can be thought of as on-site inter-orbital hopping terms, and accounting for the details of orbital structure will modify the form of the hopping between sites in Eq.~\ref{eqn:multiorb_hopping_gen}. Including both of these effects will, of course, change the structure of the non-interacting molecular orbital levels, and can further reduce the orbital symmetries. However, they will not make a dramatic qualitative change in the underlying core physics of CMIs. On the other hand, introducing spin-orbit coupling will have a dramatic effect, as it would mean spin is no longer conserved. The majority of the non-trivial GSDs encountered in the examples we have shown here are protected, not by lattice or orbital symmetries, but by spin conservation, which are thus bound to change once spin-orbit coupling is introduced. Indeed, in a realistic material, the only symmetries likely remaining will be the point group of the cluster and time-reversal symmetry. In such a scenario, only doublets (Kramers or non-Kramers) will realistically be possible (with the exception of the tetrahedral cluster, whose $T_d$ point group symmetry contains three-fold degenerate representations). This can be seen in studies of realistic $\mathrm{M_2O_{9}}$ dimer \cite{m2o9dimer} and $\mathrm{M_3O_{12}}$ trimer \cite{vjthesis} cluster materials. Beyond the cluster itself, the surrounding lattice may induce Jahn-Teller distortions and further reduce the GSD of a cluster, particularly in the rare cases encountered here in which orbital symmetries act to protect the localized degrees of freedom.

In this work, we primarily focused on clusters with odd numbers of electrons, and localized degrees of freedom with GSD $>$ 1, as these provide the ideal starting points for building non-trivial effective inter-cluster Hamiltonians. However, recent work has introduced the idea of $n$-Mott atomic limits, clusters with $n$-electrons with a unique, symmetric ground state which transforms non-trivially under lattice symmetries \cite{glennwagner}. For the single orbital case, in Fig.~\ref{fig:1orb_table}, the only example of such a scenario is the square lattice case with $n_f=4$, in which the unique ground state carries a $B_2$ irreducible representation of the $C_{4v}$ lattice symmetry (this example is also discussed in Ref.~\cite{glennwagner}). Other clusters could potentially also host non-trivial ground states \cite{vjthesis}. However, we have not found any further occurrences of such ground states amongst the parameters studied here, even amongst the even $n_f$ sectors where unique ground states are common due to the absence of Kramers degeneracies. This suggests that such Mott atomic limits may be relatively rare in practice. However, the GSD $=$ 1 region in the two-orbital tetrahedron phase diagram shown in Fig.~\ref{fig:hedron_2orbs_panel} does carry a non-trivial irreducible representation, $B$, of the $C_2$ orbital symmetry. 


Beyond traditional solid-state materials, it is also worth noting that there is significant conceptual overlap between cluster Mott insulators and the recently studied moir\'e Wigner molecules \cite{Reddy2023,Yann2023,Hong2024}. A moir\'e superlattice potential can generate an array of ``artificial atoms" which, at certain fillings, can attain a complex internal ``molecular" structure due to strong Coulomb interactions. By coupling such moir\'e molecules together within a lattice, one can construct an effective Hamiltonian describing interactions between their localized molecular degrees of freedom \cite{khalifa2025}. These moir\'e Wigner molecules, with their non-trivial internal level structure, thus play a role analogous to the clusters studied here. It would be interesting to explore in future whether some of the intuition gained here by studying simplified cluster Hamiltonians can be extended to moir\'e Wigner molecules. 

Finally, determining the potential localized degrees of freedom is just the first step in understanding the physics of CMIs. As outlined in Section \ref{model_methods}, the next step is the construction of the effective Hamiltonians governing the interactions between the localized cluster degrees of freedom, which can be computed via degenerate perturbation theory in the inter-cluster Hamiltonian $H_{CC'}$. Taking this next step will allow us to explore what kind of new, many-body physics is possible with CMIs, how they compare and contrast with the more traditional single-site Mott insulators, and help in understanding some of the outstanding experimental puzzles in CMI materials.   

\acknowledgements
We thank Maria Hermanns, Willian M.H.~Natori, Luca Peterlini and Stephen Winter for useful discussions. This research was supported in part by the National Science Foundation under Grants No.~NSF PHY-1748958 and PHY-2309135. We acknowledge support from the Deutsche Forschungsgemeinschaft (DFG, German Research Foundation) within Project-ID 277146847, SFB1238 (project C03).

\appendix

\section{Hubbard-Kanamori Hamiltonian}\label{app:HK}

In the case of multiple orbitals, it can be shown that the most general interaction term is a matrix element of the screened Coulomb interaction $V_c$, a Coulomb integral of the form \cite{interaction_integrals,coulomb_integrals}

\begin{equation}
\begin{split}
    A^{mnpq}_{ijkl} = 
\int  d\boldsymbol{r}d\boldsymbol{r'} \phi^*_{im}(\boldsymbol{r})
\phi^*_{jn}(\boldsymbol{r}')  V_c(\boldsymbol{r},\boldsymbol{r'}) 
 \phi_{kp}(\boldsymbol{r}') \phi_{lq}(\boldsymbol{r}),
\end{split}
\label{eqn:coulomb_integral}
\end{equation}
where $\phi_m(\boldsymbol{r})$ is some localized Wannier basis, $i,j,k,l$ are site indices, and $m,n,p,q$ are orbital indices. As we only consider here local interactions that decay rapidly with distance between two sites $i$ and $j$, we set all site indices as equal, and hence drop site indices in the remainder of this section. 

The on-site interaction between electrons in a single orbital, that is, the Hubbard interaction $U$, is obtained when we set  $m = n = p = q$:

\begin{equation}
    U = \int d\boldsymbol{r}d\boldsymbol{r'}|\phi_m(\boldsymbol{r})|^2V_c(\boldsymbol{r},\boldsymbol{r'})|\phi_m(\boldsymbol{r'})|^2
    \label{eqn:u_integral}
\end{equation}
A similar interaction would be an on-site term between electrons on different orbitals. If we set $m = q,
n = p$ we get

\begin{equation}
    U' = \int d\boldsymbol{r}d\boldsymbol{r'}|\phi_m(\boldsymbol{r})|^2V_c(\boldsymbol{r},\boldsymbol{r'})|\phi_{n}(\boldsymbol{r'})|^2,
    \label{eqn:up_integral}
\end{equation}
where a change of variable $p\rightarrow n$ is used. Similarly, we get two more interaction terms that are non-diagonal in occupation number, which are the $J$-interactions

\begin{equation}
\begin{split}
J_1 &= \int d\boldsymbol{r}d\boldsymbol{r'}\phi^*_m(\boldsymbol{r})\phi^*_{n}(\boldsymbol{r}')V_c(\boldsymbol{r},\boldsymbol{r'})\phi_m(\boldsymbol{r'})\phi_{n}(\boldsymbol{r})\\
J_2 &= \int d\boldsymbol{r}d\boldsymbol{r'}\phi^*_m(\boldsymbol{r})\phi^*_{m}(\boldsymbol{r'})V_c(\boldsymbol{r},\boldsymbol{r'})\phi_{n}(\boldsymbol{r'})\phi_{n}(\boldsymbol{r}).
\end{split}
\label{eqn:j_integral}
\end{equation}
If we choose $\phi_m(\boldsymbol{r})$ to be real, we get $J_1=J_2=J$. In second quantized form, the interactions (\ref{eqn:u_integral}), (\ref{eqn:up_integral}) and (\ref{eqn:j_integral}) combine to  give the Hubbard-Kanamori Hamitonian

\begin{equation}
    \begin{split}
     H_{HK}  =& \,\,U\sum_{m}n_{m\uparrow}n_{m\downarrow}+U'\sum_{m\neq n }n_{m\uparrow}n_{n\downarrow} \\
    & +(U'-J)\sum_{m\neq n,\sigma}n_{m\sigma}n_{n\sigma}  - J \sum_{m\neq n}c^{\dagger}_{m\uparrow}c^{\dagger}_{n\downarrow}c_{m\downarrow}c_{n\uparrow}\\
 & + J \sum_{m\neq n}c^{\dagger}_{m\uparrow}c^{\dagger}_{m\downarrow}c_{n\downarrow}c_{n\uparrow},
 \end{split}
 \label{eqn:hk_Hamiltonian}
\end{equation}
 where the operator $c^\dagger_{m\sigma}(c_{m\sigma})$ creates (annihilates) an electron with spin $\sigma$ in atomic orbital $m$. In the above equation, the first three terms are density-density interactions: $U$ being between opposite spins in the same orbital, $U'$ between opposite spins in  different orbitals and $U'-J$ being between parallel spins on different orbitals. The $J$-term consists of spin-flip and pair-hopping terms. The mechanisms for all terms in Eq.~\ref{eqn:hk_Hamiltonian} are illustrated in Fig.~\cref{fig:int_mech} in the main text.

 Note that Eq.~\ref{eqn:hk_Hamiltonian} is an exact description of the interactions among orbitals only when full spherical symmetry of $V_c$ and the orbitals involved is assumed. In such a case \cite{Georges_2013}
\begin{equation}
  U' = U-2J.
  \label{eqn:up_approx}
\end{equation}
For example, in the three orbital case, this relation holds if we consider a partially quenched orbital angular momentum, from $l=2$ for the entire $d$-shell, down to $l=1$. However, in the most general case, Eq.~\ref{eqn:up_approx} does not hold; there would exist terms in addition to $U,U'$ and $J$ which might not vanish by symmetry. In that case, the Hubbard-Kanamori Hamiltonian would only be approximate.

\raggedright 
\bibliography{Cluster_Mott_Insulators}

@PREAMBLE{
 "\providecommand{\noopsort}[1]{}" 
 # "\providecommand{\singleletter}[1]{#1}%" 
}

@article{khomskii_review,
	author = {Khomskii, Daniel I. and Streltsov, Sergey V.},
	date = {2021/03/10},
	doi = {10.1021/acs.chemrev.0c00579},
	isbn = {0009-2665},
	journal = {Chemical Reviews},
	journal1 = {Chemical Reviews},
	journal2 = {Chem. Rev.},
	month = {03},
	number = {5},
	pages = {2992--3030},
	publisher = {American Chemical Society},
	title = {Orbital Effects in Solids: Basics, Recent Progress, and Opportunities},
	type = {doi: 10.1021/acs.chemrev.0c00579},
	url = {https://doi.org/10.1021/acs.chemrev.0c00579},
	volume = {121},
	year = {2021},
	year1 = {2021}}

@article{triangle_CsW2O6,
	author = {Okamoto, Yoshihiko and Amano, Haruki and Katayama, Naoyuki and Sawa, Hiroshi and Niki, Kenta and Mitoka, Rikuto and Harima, Hisatomo and Hasegawa, Takumi and Ogita, Norio and Tanaka, Yu and Takigawa, Masashi and Yokoyama, Yasunori and Takehana, Kanji and Imanaka, Yasutaka and Nakamura, Yuto and Kishida, Hideo and Takenaka, Koshi},
	date = {2020/06/19},
	doi = {10.1038/s41467-020-16873-7},
	id = {Okamoto2020},
	isbn = {2041-1723},
	journal = {Nature Communications},
	number = {1},
	pages = {3144},
	title = {Regular-triangle trimer and charge order preserving the Anderson condition in the pyrochlore structure of $\mathrm{CsW}_2\mathrm{O}_6$},
	url = {https://doi.org/10.1038/s41467-020-16873-7},
	volume = {11},
	year = {2020}}

@phdthesis{vjthesis,
            year = {2025},
          author = {Vaishnavi Jayakumar},
           title = {\textnormal{Building blocks for cluster Mott insulators: from elementary models to potential realizations}},
          school = {Universit{\"a}t zu K{\"o}ln},
             url = {https://kups.ub.uni-koeln.de/78361/}
}

@article{Alaska2015,
  title = {Low-energy description of the metal-insulator transition in the rare-earth nickelates},
  author = {Subedi, Alaska and Peil, Oleg E. and Georges, Antoine},
  journal = {Phys. Rev. B},
  volume = {91},
  issue = {7},
  pages = {075128},
  numpages = {16},
  year = {2015},
  month = {Feb},
  publisher = {American Physical Society},
  doi = {10.1103/PhysRevB.91.075128},
  url = {https://link.aps.org/doi/10.1103/PhysRevB.91.075128}
}

@article{Ryee2021,
  title = {Hund Physics Landscape of Two-Orbital Systems},
  author = {Ryee, Siheon and Han, Myung Joon and Choi, Sangkook},
  journal = {Phys. Rev. Lett.},
  volume = {126},
  issue = {20},
  pages = {206401},
  numpages = {7},
  year = {2021},
  month = {May},
  publisher = {American Physical Society},
  doi = {10.1103/PhysRevLett.126.206401},
  url = {https://link.aps.org/doi/10.1103/PhysRevLett.126.206401}
}

@article{Isidori2019,
  title = {Charge Disproportionation, Mixed Valence, and Janus Effect in Multiorbital Systems: A Tale of Two Insulators},
  author = {Isidori, Aldo and Berovi\ifmmode \acute{c}\else \'{c}\fi{}, Maja and Fanfarillo, Laura and de' Medici, Luca and Fabrizio, Michele and Capone, Massimo},
  journal = {Phys. Rev. Lett.},
  volume = {122},
  issue = {18},
  pages = {186401},
  numpages = {6},
  year = {2019},
  month = {May},
  publisher = {American Physical Society},
  doi = {10.1103/PhysRevLett.122.186401},
  url = {https://link.aps.org/doi/10.1103/PhysRevLett.122.186401}
}

@article{m2o9dimer,
  title = {Soft and anisotropic local moments in $4d$ and $5d$ mixed-valence {$\mathrm{M}_2\mathrm{O}_9$} dimers},
  author = {Li, Ying and Tsirlin, Alexander A. and Dey, Tusharkanti and Gegenwart, Philipp and Valent\'{\i}, Roser and Winter, Stephen M.},
  journal = {Phys. Rev. B},
  volume = {102},
  issue = {23},
  pages = {235142},
  numpages = {12},
  year = {2020},
  month = {Dec},
  publisher = {American Physical Society},
  doi = {10.1103/PhysRevB.102.235142},
  url = {https://link.aps.org/doi/10.1103/PhysRevB.102.235142}
}

@article{Georges_2013,
	doi = {10.1146/annurev-conmatphys-020911-125045},
	url = { 	
https://doi.org/10.1146/annurev-conmatphys-020911-125045},
	year = 2013,
	month = {apr},
	publisher = {Annual Reviews},
	volume = {4},
	number = {1},
	pages = {137--178},
	author = {Antoine Georges and Luca de{\textquotesingle} Medici and Jernej Mravlje},
	title = {Strong Correlations from {H}und's Coupling},
	journal = {Annual Review of Condensed Matter Physics}
}

@article{Qiang_Ba3LaRu2O9,
  title = {Realization of the orbital-selective {M}ott state at the molecular level in $\mathrm{Ba}_{3}\mathrm{LaRu}_{2}\mathrm{O}_{9}$},
  author = {Chen, Q. and Verrier, A. and Ziat, D. and Clune, A. J. and Rouane, R. and Bazier-Matte, X. and Wang, G. and Calder, S. and Taddei, K. M. and Cruz, C. R. dela and Kolesnikov, A. I. and Ma, J. and Cheng, J.-G. and Liu, Z. and Quilliam, J. A. and Musfeldt, J. L. and Zhou, H. D. and Aczel, A. A.},
  journal = {Phys. Rev. Mater.},
  volume = {4},
  issue = {6},
  pages = {064409},
  numpages = {12},
  year = {2020},
  month = {Jun},
  publisher = {American Physical Society},
  doi = {10.1103/PhysRevMaterials.4.064409},
  url = {https://link.aps.org/doi/10.1103/PhysRevMaterials.4.064409}
}

@article{Qiang_Momagnets,
  title = {Magnetic order and spin liquid behavior in ${[{\mathrm{Mo}}_{3}]}^{11+}$ molecular magnets},
  author = {Chen, Q. and Sinclair, R. and Akbari-Sharbaf, A. and Huang, Q. and Dun, Z. and Choi, E. S. and Mourigal, M. and Verrier, A. and Rouane, R. and Bazier-Matte, X. and Quilliam, J. A. and Aczel, A. A. and Zhou, H. D.},
  journal = {Phys. Rev. Mater.},
  volume = {6},
  issue = {4},
  pages = {044414},
  numpages = {12},
  year = {2022},
  month = {Apr},
  publisher = {American Physical Society},
  doi = {10.1103/PhysRevMaterials.6.044414},
  url = {https://link.aps.org/doi/10.1103/PhysRevMaterials.6.044414}
}

@article{chargeorder_Ba3NaRu2O9,
  title = {Charge Order at the Frontier between the Molecular and Solid States in $\mathrm{Ba}_{3}\mathrm{NaRu}_{2}\mathbf{O}_{9}$},
  author = {Kimber, Simon A. J. and Senn, Mark S. and Fratini, Simone and Wu, Hua and Hill, Adrian H. and Manuel, Pascal and Attfield, J. Paul and Argyriou, Dimitri N. and Henry, Paul. F.},
  journal = {Phys. Rev. Lett.},
  volume = {108},
  issue = {21},
  pages = {217205},
  numpages = {5},
  year = {2012},
  month = {May},
  publisher = {American Physical Society},
  doi = {10.1103/PhysRevLett.108.217205},
  url = {https://link.aps.org/doi/10.1103/PhysRevLett.108.217205}
}

@article{lacunarspinel_sc,
	author = {Park, Moon Jip and Sim, GiBaik and Jeong, Min Yong and Mishra, Archana and Han, Myung Joon and Lee, SungBin},
	date = {2020/06/24},
	doi = {10.1038/s41535-020-0246-0},
	id = {Park2020},
	isbn = {2397-4648},
	journal = {npj Quantum Materials},
	number = {1},
	pages = {41},
	title = {Pressure-induced topological superconductivity in the spin--orbit {M}ott insulator $\mathrm{GaTa}_4\mathrm{Se}_8$},
	url = {https://doi.org/10.1038/s41535-020-0246-0},
	volume = {5},
	year = {2020}}

@article{glennwagner,
  title = {Interacting topological quantum chemistry of Mott atomic limits},
  author = {Soldini, Martina O. and Astrakhantsev, Nikita and Iraola, Mikel and Tiwari, Apoorv and Fischer, Mark H. and Valent\'{\i}, Roser and Vergniory, Maia G. and Wagner, Glenn and Neupert, Titus},
  journal = {Phys. Rev. B},
  volume = {107},
  issue = {24},
  pages = {245145},
  numpages = {22},
  year = {2023},
  month = {Jun},
  publisher = {American Physical Society},
  doi = {10.1103/PhysRevB.107.245145},
  url = {https://link.aps.org/doi/10.1103/PhysRevB.107.245145}
}

@article{Cava_dimerkagome,
	author = {Devlin, Kasey P. and Chen, Tong and Xie, Weiwei and Broholm, Collin L. and Cava, Robert J.},
	date = {2023/05/23},
	doi = {10.1021/acsaelm.2c01501},
	journal = {ACS Applied Electronic Materials},
	journal1 = {ACS Applied Electronic Materials},
	journal2 = {ACS Appl. Electron. Mater.},
	month = {05},
	number = {5},
	pages = {2482--2486},
	publisher = {American Chemical Society},
	title = {Magnetic Material with a Kagome Net of Dimers},
	type = {doi: 10.1021/acsaelm.2c01501},
	url = {https://doi.org/10.1021/acsaelm.2c01501},
	volume = {5},
	year = {2023},
	year1 = {2023}}

@article{Gangchen_clustertransition,
  title = {Clusterization transition between cluster {M}ott insulators on a breathing kagome lattice},
  author = {Yao, Xu-Ping and Zhang, Xiao-Tian and Kim, Yong Baek and Wang, Xiaoqun and Chen, Gang},
  journal = {Phys. Rev. Res.},
  volume = {2},
  issue = {4},
  pages = {043424},
  numpages = {9},
  year = {2020},
  month = {Dec},
  publisher = {American Physical Society},
  doi = {10.1103/PhysRevResearch.2.043424},
  url = {https://link.aps.org/doi/10.1103/PhysRevResearch.2.043424}
}

@article{spinel_bondorder,
  title = {Bond ordering and molecular spin-orbital fluctuations in the cluster {M}ott insulator $\mathrm{GaTa}_{4}\mathrm{Se}_{8}$},
  author = {Yang, Tsung-Han and Kawamoto, S. and Higo, Tomoya and Wang, SuYin Grass and Stone, M. B. and Neuefeind, Joerg and Ruff, Jacob P. C. and Abeykoon, A. M. Milinda and Chen, Yu-Sheng and Nakatsuji, S. and Plumb, K. W.},
  journal = {Phys. Rev. Res.},
  volume = {4},
  issue = {3},
  pages = {033123},
  numpages = {16},
  year = {2022},
  month = {Aug},
  publisher = {American Physical Society},
  doi = {10.1103/PhysRevResearch.4.033123},
  url = {https://link.aps.org/doi/10.1103/PhysRevResearch.4.033123}
}

@article{gangchen_pyrochlore,
	url = {https://doi.org/10.1103%2Fphysrevlett.113.197202},
	year = 2014,
	month = {nov},
	publisher = {American Physical Society ({APS})},
	volume = {113},
	number = {19},
	author = {Gang Chen and Hae-Young Kee and Yong Baek Kim},
	title = {Fractionalized Charge Excitations in a Spin Liquid on Partially Filled Pyrochlore Lattices},
	journal = {Physical Review Letters}
}

@article{Cava_structural,
	author = {Guo, Shu and Mitchell Warden, Hillary E. and Cava, R. J.},
	date = {2022/07/04},
	doi = {10.1021/acs.inorgchem.2c00957},
	isbn = {0020-1669},
	journal = {Inorganic Chemistry},
	journal1 = {Inorganic Chemistry},
	journal2 = {Inorg. Chem.},
	month = {07},
	number = {26},
	pages = {10043--10050},
	publisher = {American Chemical Society},
	title = {Structural Diversity in Oxoiridates with 1{D} $\mathrm{Ir}_n\mathrm{O}_{3(n+1)}$ Chain Fragments and Flat Bands},
	type = {doi: 10.1021/acs.inorgchem.2c00957},
	url = {https://doi.org/10.1021/acs.inorgchem.2c00957},
	volume = {61},
	year = {2022},
	year1 = {2022}}

@article{Nb3X8_triangular_clusters,
	author = {Pasco, Christopher M. and El Baggari, Ismail and Bianco, Elisabeth and Kourkoutis, Lena F. and McQueen, Tyrel M.},
	date = {2019/08/27},
	doi = {10.1021/acsnano.9b04392},
	isbn = {1936-0851},
	journal = {ACS Nano},
	journal1 = {ACS Nano},
	journal2 = {ACS Nano},
	month = {08},
	number = {8},
	pages = {9457--9463},
	publisher = {American Chemical Society},
	title = {Tunable Magnetic Transition to a Singlet Ground State in a 2{D} van der {W}aals Layered Trimerized Kagom{\'e} Magnet},
	type = {doi: 10.1021/acsnano.9b04392},
	url = {https://doi.org/10.1021/acsnano.9b04392},
	volume = {13},
	year = {2019},
	year1 = {2019}}

@article{winter_kitaev,
	author = {Winter, Stephen M and Tsirlin, Alexander A and Daghofer, Maria and van den Brink, Jeroen and Singh, Yogesh and Gegenwart, Philipp and Valent{\'\i}, Roser},
	date = {2017/11/13},
	doi = {10.1088/1361-648X/aa8cf5},
	isbn = {0953-8984},
	journal = {Journal of Physics: Condensed Matter},
	number = {49},
	pages = {493002},
	publisher = {IOP Publishing},
	title = {Models and materials for generalized {K}itaev magnetism},
	url = {https://dx.doi.org/10.1088/1361-648X/aa8cf5},
	volume = {29},
	year = {2017}}

@article{RIXS_Ba3InIr2O9,
  title = {Quasimolecular electronic structure of the spin-liquid candidate $\mathrm{Ba}_{3}\mathrm{InIr}_{2}\mathrm{O}_{9}$},
  author = {Revelli, A. and Moretti Sala, M. and Monaco, G. and Magnaterra, M. and Attig, J. and Peterlini, L. and Dey, T. and Tsirlin, A. A. and Gegenwart, P. and Fr\"ohlich, T. and Braden, M. and Grams, C. and Hemberger, J. and Becker, P. and van Loosdrecht, P. H. M. and Khomskii, D. I. and van den Brink, J. and Hermanns, M. and Gr\"uninger, M.},
  journal = {Phys. Rev. B},
  volume = {106},
  issue = {15},
  pages = {155107},
  numpages = {14},
  year = {2022},
  month = {Oct},
  publisher = {American Physical Society},
  doi = {10.1103/PhysRevB.106.155107},
  url = {https://link.aps.org/doi/10.1103/PhysRevB.106.155107}
}

@article{coulomb_integrals,
	author = {Noce, Canio and Romano, Alfonso},
	date = {2014/04/01},
	doi = {https://doi.org/10.1002/pssb.201350148},
	isbn = {0370-1972},
	journal = {physica status solidi (b)},
	journal1 = {physica status solidi (b)},
	journal2 = {physica status solidi (b)},
	journal3 = {Phys. Status Solidi B},
	month = {2023/09/27},
	number = {4},
	pages = {907--911},
	publisher = {John Wiley \& Sons, Ltd},
	title = {Rotationally invariant parametrization of {C}oulomb interactions in multi-orbital {H}ubbard models},
	url = {https://doi.org/10.1002/pssb.201350148},
	volume = {251},
	year = {2014},
	year1 = {2014}}

@article{kitaevmaterials_ciaran,
	author = {Trebst, Simon and Hickey, Ciar{\'a}n},
	date = {2022/03/04/},
	doi = {https://doi.org/10.1016/j.physrep.2021.11.003},
	isbn = {0370-1573},
	journal = {Physics Reports},
	journal1 = {Kitaev materials},
	pages = {1--37},
	title = {{K}itaev materials},
	url = {https://www.sciencedirect.com/science/article/pii/S0370157321004051},
	volume = {950},
	year = {2022}}

@article{bjkim,
  title = {Novel ${J}_{\mathrm{eff}}=1/2$ {M}ott State Induced by Relativistic Spin-Orbit Coupling in $\mathrm{Sr}_{2}\mathrm{IrO}_{4}$},
  author = {Kim, B. J. and Jin, Hosub and Moon, S. J. and Kim, J.-Y. and Park, B.-G. and Leem, C. S. and Yu, Jaejun and Noh, T. W. and Kim, C. and Oh, S.-J. and Park, J.-H. and Durairaj, V. and Cao, G. and Rotenberg, E.},
  journal = {Phys. Rev. Lett.},
  volume = {101},
  issue = {7},
  pages = {076402},
  numpages = {4},
  year = {2008},
  month = {Aug},
  publisher = {American Physical Society},
  doi = {10.1103/PhysRevLett.101.076402},
  url = {https://link.aps.org/doi/10.1103/PhysRevLett.101.076402}
}

@article{Qiang_Ba3CeRu2O9,
	author = {Chen, Qiang and Fan, Shiyu and Taddei, Keith M. and Stone, Matthew B. and Kolesnikov, Alexander I. and Cheng, Jinguang and Musfeldt, Janice L. and Zhou, Haidong and Aczel, Adam A.},
	date = {2019/06/26},
	doi = {10.1021/jacs.9b03389},
	isbn = {0002-7863},
	journal = {Journal of the American Chemical Society},
	journal1 = {Journal of the American Chemical Society},
	journal2 = {J. Am. Chem. Soc.},
	month = {06},
	number = {25},
	pages = {9928--9936},
	publisher = {American Chemical Society},
	title = {Large Positive Zero-Field Splitting in the Cluster Magnet $\mathrm{Ba}_3\mathrm{CeRu}_2\mathrm{O}9$},
	type = {doi: 10.1021/jacs.9b03389},
	url = {https://doi.org/10.1021/jacs.9b03389},
	volume = {141},
	year = {2019},
	year1 = {2019}}

@article{spin0spin1,
  title = {Spin-0 {M}ott insulator to metal to spin-1 {M}ott insulator transition in the single-orbital {H}ubbard model on the decorated honeycomb lattice},
  author = {Nourse, H. L. and McKenzie, Ross H. and Powell, B. J.},
  journal = {Phys. Rev. B},
  volume = {104},
  issue = {7},
  pages = {075104},
  numpages = {14},
  year = {2021},
  month = {Aug},
  publisher = {American Physical Society},
  doi = {10.1103/PhysRevB.104.075104},
  url = {https://link.aps.org/doi/10.1103/PhysRevB.104.075104}
}

@article{Cava,
	doi = {10.1021/acs.chemrev.0c00622},
	url = {https://doi.org/10.1021\%2Facs.chemrev.0c00622},
	year = 2020,
	month = {sep},
	publisher = {American Chemical Society ({ACS})},
	volume = {121},
	number = {5},
	pages = {2935--2965},
	author = {Loi T. Nguyen and R. J. Cava},
	title = {Hexagonal Perovskites as Quantum Materials},
	journal = {Chemical Reviews}
}

@phdthesis{interaction_integrals,
    title={\normalfont{Exploring {H}und’s correlated metals: charge instabilities and effect of selective interactions}},
    school   = {SISSA},
     author={Berovic, Maja},
    url = {https://hdl.handle.net/20.500.11767/84088},
  author={Berovic, Maja},
  year={2018},
  publisher={SISSA} 
}

@article{honeycomb_LiZn2Mo3O8,
  title = {Emergent Honeycomb Lattice in $\mathrm{LiZn}_{2}\mathrm{Mo}_{3}\mathrm{O}_{8}$},
  author = {Flint, Rebecca and Lee, Patrick A.},
  journal = {Phys. Rev. Lett.},
  volume = {111},
  issue = {21},
  pages = {217201},
  numpages = {5},
  year = {2013},
  month = {Nov},
  publisher = {American Physical Society},
  doi = {10.1103/PhysRevLett.111.217201},
  url = {https://link.aps.org/doi/10.1103/PhysRevLett.111.217201}
}

@article{Gangchen_CMI_CurieWeiss,
  title = {Cluster {M}ott insulators and two Curie-Weiss regimes on an anisotropic kagome lattice},
  author = {Chen, Gang and Kee, Hae-Young and Kim, Yong Baek},
  journal = {Phys. Rev. B},
  volume = {93},
  issue = {24},
  pages = {245134},
  numpages = {12},
  year = {2016},
  month = {Jun},
  publisher = {American Physical Society},
  doi = {10.1103/PhysRevB.93.245134},
  url = {https://link.aps.org/doi/10.1103/PhysRevB.93.245134}
}

@article{Gangchen_kagome,
  title = {Emergent orbitals in the cluster {M}ott insulator on a breathing kagome lattice},
  author = {Chen, Gang and Lee, Patrick A.},
  journal = {Phys. Rev. B},
  volume = {97},
  issue = {3},
  pages = {035124},
  numpages = {12},
  year = {2018},
  month = {Jan},
  publisher = {American Physical Society},
  doi = {10.1103/PhysRevB.97.035124},
  url = {https://link.aps.org/doi/10.1103/PhysRevB.97.035124}
}

@article{clustermagnet_vanderwaals,
	doi = {10.1039/c6qi00470a},
	url = {https://doi.org/10.1039%2Fc6qi00470a},
	year = 2017,
	publisher = {Royal Society of Chemistry ({RSC})},
	volume = {4},
	number = {3},
	pages = {481--490},
	author = {John P. Sheckelton and Kemp W. Plumb and Benjamin A. Trump and Collin L. Broholm and Tyrel M. McQueen},
	title = {Rearrangement of van der {W}aals stacking and formation of a singlet state at $\mathrm{T} = 90 \mathrm{ K}$ in a cluster magnet},
	journal = {Inorganic Chemistry Frontiers}
}

@article{qmagnetism_milestones,
	author = {Vasiliev, Alexander and Volkova, Olga and Zvereva, Elena and Markina, Maria},
	date = {2018/03/28},
	doi = {10.1038/s41535-018-0090-7},
	id = {Vasiliev2018},
	isbn = {2397-4648},
	journal = {npj Quantum Materials},
	number = {1},
	pages = {18},
	title = {Milestones of low-{D} quantum magnetism},
	url = {https://doi.org/10.1038/s41535-018-0090-7},
	volume = {3},
	year = {2018}}

@article{mott_jackeli,
  title = {{M}ott Insulators in the Strong Spin-Orbit Coupling Limit: From {H}eisenberg to a Quantum Compass and {K}itaev Models},
  author = {Jackeli, G. and Khaliullin, G.},
  journal = {Phys. Rev. Lett.},
  volume = {102},
  issue = {1},
  pages = {017205},
  numpages = {4},
  year = {2009},
  month = {Jan},
  publisher = {American Physical Society},
  doi = {10.1103/PhysRevLett.102.017205},
  url = {https://link.aps.org/doi/10.1103/PhysRevLett.102.017205}
}

@article{cuprate_rmp,
  title = {Doping a {M}ott insulator: Physics of high-temperature superconductivity},
  author = {Lee, Patrick A. and Nagaosa, Naoto and Wen, Xiao-Gang},
  journal = {Rev. Mod. Phys.},
  volume = {78},
  issue = {1},
  pages = {17--85},
  numpages = {0},
  year = {2006},
  month = {Jan},
  publisher = {American Physical Society},
  doi = {10.1103/RevModPhys.78.17},
  url = {https://link.aps.org/doi/10.1103/RevModPhys.78.17}
}

@article{heptamers,
  title = {Spontaneous Formation of Vanadium ``Molecules'' in a Geometrically Frustrated Crystal: $\mathrm{AlV}_{2}\mathrm{O}_{4}$},
  author = {Horibe, Y. and Shingu, M. and Kurushima, K. and Ishibashi, H. and Ikeda, N. and Kato, K. and Motome, Y. and Furukawa, N. and Mori, S. and Katsufuji, T.},
  journal = {Phys. Rev. Lett.},
  volume = {96},
  issue = {8},
  pages = {086406},
  numpages = {4},
  year = {2006},
  month = {Mar},
  publisher = {American Physical Society},
  doi = {10.1103/PhysRevLett.96.086406},
  url = {https://link.aps.org/doi/10.1103/PhysRevLett.96.086406}
}

@article{LiVO2,
  title = {Orbital Ordering in a Two-Dimensional Triangular Lattice},
  author = {Pen, H. F. and van den Brink, J. and Khomskii, D. I. and Sawatzky, G. A.},
  journal = {Phys. Rev. Lett.},
  volume = {78},
  issue = {7},
  pages = {1323--1326},
  numpages = {0},
  year = {1997},
  month = {Feb},
  publisher = {American Physical Society},
  doi = {10.1103/PhysRevLett.78.1323},
  url = {https://link.aps.org/doi/10.1103/PhysRevLett.78.1323}
}

@article{pyroxene,
  title = {Orbital and spin interplay in spin-gap formation in pyroxene $\mathrm{ATiSi}_{2}\mathrm{O}_{6}$ $(A=\mathrm{Na},\mathrm{Li})$},
  author = {Hikihara, Toshiya and Motome, Yukitoshi},
  journal = {Phys. Rev. B},
  volume = {70},
  issue = {21},
  pages = {214404},
  numpages = {12},
  year = {2004},
  month = {Dec},
  publisher = {American Physical Society},
  doi = {10.1103/PhysRevB.70.214404},
  url = {https://link.aps.org/doi/10.1103/PhysRevB.70.214404}
}

@article{Nb3Cl8_phasetrans,
	author = {Haraguchi, Yuya and Michioka, Chishiro and Ishikawa, Manabu and Nakano, Yoshiaki and Yamochi, Hideki and Ueda, Hiroaki and Yoshimura, Kazuyoshi},
	date = {2017/03/20},
	doi = {10.1021/acs.inorgchem.6b03028},
	isbn = {0020-1669},
	journal = {Inorganic Chemistry},
	journal1 = {Inorganic Chemistry},
	journal2 = {Inorg. Chem.},
	month = {03},
	number = {6},
	pages = {3483--3488},
	publisher = {American Chemical Society},
	title = {Magnetic--Nonmagnetic Phase Transition with Interlayer Charge Disproportionation of $\mathrm{Nb}_3$ Trimers in the Cluster Compound $\mathrm{Nb}_3\mathrm{Cl}_8$},
	type = {doi: 10.1021/acs.inorgchem.6b03028},
	url = {https://doi.org/10.1021/acs.inorgchem.6b03028},
	volume = {56},
	year = {2017},
	year1 = {2017}}

@article{nikolaev,
	date = {2021/03/11},
	doi = {10.1038/s41535-021-00316-7},
	id = {Nikolaev2021},
	isbn = {2397-4648},
 author = {Nikolaev, S. A. and Solovyev, I. V. and Streltsov, S. V.},
	journal = {npj Quantum Materials},
	number = {1},
	pages = {25},
	title = {Quantum spin liquid and cluster {M}ott insulator phases in the {$\mathrm{Mo}_3\mathrm{O}_8$} magnets},
	url = {https://doi.org/10.1038/s41535-021-00316-7},
	volume = {6},
	year = {2021}}

@article{Mo3o8_valencebond,
	author = {Sheckelton, J. P. and Neilson, J. R. and Soltan, D. G. and McQueen, T. M.},
	date = {2012/06/01},
	doi = {10.1038/nmat3329},
	id = {Sheckelton2012},
	isbn = {1476-4660},
	journal = {Nature Materials},
	number = {6},
	pages = {493--496},
	title = {Possible valence-bond condensation in the frustrated cluster magnet $\mathrm{LiZn}_2\mathrm{Mo}_3\mathrm{O}_8$},
	url = {https://doi.org/10.1038/nmat3329},
	volume = {11},
	year = {2012}}

@article{Khomskii_SOCandHund,
  title = {Suppression of magnetism in $\mathrm{Ba}_{5}\mathrm{AlIr}_{2}\mathrm{O}_{11}$: Interplay of $\mathrm{H}$und's coupling, molecular orbitals, and spin-orbit interaction},
  author = {Streltsov, Sergey V. and Cao, Gang and Khomskii, Daniel I.},
  journal = {Phys. Rev. B},
  volume = {96},
  issue = {1},
  pages = {014434},
  numpages = {5},
  year = {2017},
  month = {Jul},
  publisher = {American Physical Society},
  doi = {10.1103/PhysRevB.96.014434},
  url = {https://link.aps.org/doi/10.1103/PhysRevB.96.014434}
}

@article{nb3cl8_flatbands,
	author = {Hu, Jiayu and Zhang, Xuefeng and Hu, Cong and Sun, Jian and Wang, Xiaoqun and Lin, Hai-Qing and Li, Gang},
	date = {2023/07/11},
	doi = {10.1038/s42005-023-01292-z},
	id = {Hu2023},
	isbn = {2399-3650},
	journal = {Communications Physics},
	number = {1},
	pages = {172},
	title = {Correlated flat bands and quantum spin liquid state in a cluster Mott insulator},
	url = {https://doi.org/10.1038/s42005-023-01292-z},
	volume = {6},
	year = {2023}}

@article{Ba3ZnIr2O9,
  title = {Origin of the Spin-Orbital Liquid State in a Nearly {$J=0$} Iridate {$\mathrm{Ba}_3\mathrm{ZnIr}_2\mathrm{O}_9$}},
  author = {Nag, Abhishek and Middey, S. and Bhowal, Sayantika and Panda, S. K. and Mathieu, Roland and Orain, J. C. and Bert, F. and Mendels, P. and Freeman, P. G. and Mansson, M. and Ronnow, H. M. and Telling, M. and Biswas, P. K. and Sheptyakov, D. and Kaushik, S. D. and Siruguri, Vasudeva and Meneghini, Carlo and Sarma, D. D. and Dasgupta, Indra and Ray, Sugata},
  journal = {Phys. Rev. Lett.},
  volume = {116},
  issue = {9},
  pages = {097205},
  numpages = {5},
  year = {2016},
  month = {Mar},
  publisher = {American Physical Society},
  doi = {10.1103/PhysRevLett.116.097205},
  url = {https://link.aps.org/doi/10.1103/PhysRevLett.116.097205}
}

@article{RIXS_M2O9_disorder,
	author = {Magnaterra, M. and Moretti Sala, M. and Monaco, G. and Becker, P. and Hermanns, M. and Warzanowski, P. and Lorenz, T. and Khomskii, D. I. and van Loosdrecht, P. H. M. and van den Brink, J. and Gr{\"u}ninger, M.},
	date = {2023/03/10/},
	day = {10},
	doi = {10.1103/PhysRevResearch.5.013167},
	id = {10.1103/PhysRevResearch.5.013167},
	j1 = {PRRESEARCH},
	journal = {Physical Review Research},
	journal1 = {Phys. Rev. Res.},
	month = {03},
	number = {1},
	pages = {013167--},
	publisher = {American Physical Society},
	title = {$\mathrm{RIXS}$ interferometry and the role of disorder in the quantum magnet $\mathrm{Ba}_3\mathrm{Ti}_{\mathrm{3-x}}\mathrm{Ir}_{\mathrm{x}}\mathrm{O}_9$},
	url = {https://link.aps.org/doi/10.1103/PhysRevResearch.5.013167},
	volume = {5},
	year = {2023}}

@article{GaTa4Se8,
	author = {Jeong, Min Yong and Chang, Seo Hyoung and Kim, Beom Hyun and Sim, Jae-Hoon and Said, Ayman and Casa, Diego and Gog, Thomas and Janod, Etienne and Cario, Laurent and Yunoki, Seiji and Han, Myung Joon and Kim, Jungho},
	date = {2017/10/04},
	doi = {10.1038/s41467-017-00841-9},
	id = {Jeong2017},
	isbn = {2041-1723},
	journal = {Nature Communications},
	number = {1},
	pages = {782},
	title = {Direct experimental observation of the molecular {$J_\mathrm{eff}=3/2$} ground state in the lacunar spinel {$\mathrm{GaTa}_4\mathrm{Se}_8$}},
	url = {https://doi.org/10.1038/s41467-017-00841-9},
	volume = {8},
	year = {2017}}

@article{Ba4spinliq,
	author = {Thakur, Gohil S. and Chattopadhyay, Sumanta and Doert, Thomas and Herrmannsd{\"o}rfer, T. and Felser, Claudia},
	date = {2020/05/06},
	doi = {10.1021/acs.cgd.0c00262},
	isbn = {1528-7483},
	journal = {Crystal Growth \& Design},
	journal1 = {Crystal Growth \& Design},
	journal2 = {Crystal Growth \& Design},
	month = {05},
	number = {5},
	pages = {2871--2876},
	publisher = {American Chemical Society},
	title = {Crystal Growth of Spin-frustrated {$\mathrm{Ba}_4\mathrm{Nb}_{0.8}\mathrm{Ir}_{3.2}\mathrm{O}_{12}$}: A Possible Spin Liquid Material},
	type = {doi: 10.1021/acs.cgd.0c00262},
	url = {https://doi.org/10.1021/acs.cgd.0c00262},
	volume = {20},
	year = {2020},
	year1 = {2020}}

@phdthesis{casimir,
  title={\normalfont{Correlated Materials - Models \& Methods}},
  author={Hugo U. R. Strand},
  year={2013},
  school   = {University of Gothenburg},
  url={https://gupea.ub.gu.se/handle/2077/32118}
}

@article{herbertsmithite,
  title = {Colloquium: Herbertsmithite and the search for the quantum spin liquid},
  author = {Norman, M. R.},
  journal = {Rev. Mod. Phys.},
  volume = {88},
  issue = {4},
  pages = {041002},
  numpages = {14},
  year = {2016},
  month = {Dec},
  publisher = {American Physical Society},
  doi = {10.1103/RevModPhys.88.041002},
  url = {https://link.aps.org/doi/10.1103/RevModPhys.88.041002}
}

@article{qsl_review,
	author = {Savary, Lucile and Balents, Leon},
	date = {2016/11/08},
	doi = {10.1088/0034-4885/80/1/016502},
	isbn = {0034-4885},
	journal = {Reports on Progress in Physics},
	number = {1},
	pages = {016502},
	publisher = {IOP Publishing},
	title = {Quantum spin liquids: a review},
	url = {https://dx.doi.org/10.1088/0034-4885/80/1/016502},
	volume = {80},
	year = {2017}}

@article{organic_mott,
	author = {Kanoda, Kazushi and Kato, Reizo},
	date = {2011/03/01},
	doi = {10.1146/annurev-conmatphys-062910-140521},
	isbn = {1947-5454},
	journal = {Annual Review of Condensed Matter Physics},
	journal1 = {Annual Review of Condensed Matter Physics},
	journal2 = {Annu. Rev. Condens. Matter Phys.},
	month = {2023/10/01},
	number = {1},
	pages = {167--188},
	publisher = {Annual Reviews},
	title = {Mott Physics in Organic Conductors with Triangular Lattices},
	type = {doi: 10.1146/annurev-conmatphys-062910-140521},
	url = {https://doi.org/10.1146/annurev-conmatphys-062910-140521},
	volume = {2},
	year = {2011},
	year1 = {2011}}

@inbook{sachdev_mott,
	address = {Berlin, Heidelberg},
	author = {Sachdev, Subir},
	booktitle = {Quantum Magnetism},
	date = {2004//},
	doi = {10.1007/BFb0119599},
	editor = {Schollw{\"o}ck, Ulrich and Richter, Johannes and Farnell, Damian J. J. and Bishop, Raymod F.},
	id = {Sachdev2004},
	isbn = {978-3-540-40066-0},
	pages = {381--432},
	publisher = {Springer Berlin Heidelberg},
	title = {Quantum phases and phase transitions of {M}ott insulators},
	url = {https://doi.org/10.1007/BFb0119599},
	year = {2004}}

@Dataset{cmi_data,
  author    = {Jayakumar, Vaishnavi and Hickey, Ciarán},
  title     = {{ Elementary Building Blocks for Cluster Mott Insulators [Data set]}},
  month     = oct,
  year      = {2023},
  publisher = {Zenodo},
  url       = {https://zenodo.org/records/10041370},
}

@article{Hong2024,
author = {Hongyuan Li  and Ziyu Xiang  and Aidan P. Reddy  and Trithep Devakul  and Renee Sailus  and Rounak Banerjee  and Takashi Taniguchi  and Kenji Watanabe  and Sefaattin Tongay  and Alex Zettl  and Liang Fu  and Michael F. Crommie  and Feng Wang },
title = {Wigner molecular crystals from multielectron moiré artificial atoms},
journal = {Science},
volume = {385},
number = {6704},
pages = {86-91},
year = {2024},
doi = {10.1126/science.adk1348},
URL = {https://www.science.org/doi/abs/10.1126/science.adk1348}
}

@article{Yann2023,
  title = {Quantum Wigner molecules in moir\'e materials},
  author = {Yannouleas, Constantine and Landman, Uzi},
  journal = {Phys. Rev. B},
  volume = {108},
  issue = {12},
  pages = {L121411},
  numpages = {7},
  year = {2023},
  month = {Sep},
  publisher = {American Physical Society},
  doi = {10.1103/PhysRevB.108.L121411},
  url = {https://link.aps.org/doi/10.1103/PhysRevB.108.L121411}
}

@article{Reddy2023,
  title = {Artificial Atoms, Wigner Molecules, and an Emergent Kagome Lattice in Semiconductor Moir\'e Superlattices},
  author = {Reddy, Aidan P. and Devakul, Trithep and Fu, Liang},
  journal = {Phys. Rev. Lett.},
  volume = {131},
  issue = {24},
  pages = {246501},
  numpages = {6},
  year = {2023},
  month = {Dec},
  publisher = {American Physical Society},
  doi = {10.1103/PhysRevLett.131.246501},
  url = {https://link.aps.org/doi/10.1103/PhysRevLett.131.246501}
}

@misc{khalifa2025,
      title={Spin-orbital magnetism in moir\'e Wigner molecules}, 
      author={Ahmed Khalifa and Rokas Veitas and Francisco Machado and Shubhayu Chatterjee},
      year={2025},
      eprint={2507.06307},
      archivePrefix={arXiv},
      primaryClass={cond-mat.str-el},
      url={https://arxiv.org/abs/2507.06307}, 
}

@article{Wigner1978,
  title = {Generalized Wigner lattices in one dimension and some applications to tetracyanoquinodimethane (TCNQ) salts},
  author = {Hubbard, J.},
  journal = {Phys. Rev. B},
  volume = {17},
  issue = {2},
  pages = {494--505},
  numpages = {0},
  year = {1978},
  month = {Jan},
  publisher = {American Physical Society},
  doi = {10.1103/PhysRevB.17.494},
  url = {https://link.aps.org/doi/10.1103/PhysRevB.17.494}
}

@article{heavyfermiontrimer,
  title = {Transition between Heavy-Fermion-Strange-Metal and Quantum Spin Liquid in a $4d$-Electron Trimer Lattice},
  author = {Zhao, Hengdi and Zhang, Yu and Schlottmann, Pedro and Nandkishore, Rahul and DeLong, Lance E. and Cao, Gang},
  journal = {Phys. Rev. Lett.},
  volume = {132},
  issue = {22},
  pages = {226503},
  numpages = {7},
  year = {2024},
  month = {May},
  publisher = {American Physical Society},
  doi = {10.1103/PhysRevLett.132.226503},
  url = {https://link.aps.org/doi/10.1103/PhysRevLett.132.226503}
}

\end{document}